\documentclass[aps,pra,superscriptaddress,twocolumn,amsmath,amssymb]{revtex4-1}
\pdfoutput=1

\usepackage{microtype}

\usepackage{braket}

\usepackage{amsmath}

\usepackage{float}

\usepackage{graphics}
\usepackage{caption}
\usepackage{subcaption}

\usepackage{tikz}
\usetikzlibrary{decorations.markings}
\usetikzlibrary{shapes,backgrounds}

\newcommand{\e}{\mathrm{e}}
\renewcommand{\iota}{\mathrm{i}}

\usepackage[colorlinks]{hyperref}

\begin{document}

\title{Quantum-optical tests of Planck-scale physics}

\author{Shreya P. Kumar}
\address{Institute of Theoretical Physics and Center for Integrated Quantum Science and Technology (IQST), Albert-Einstein-Allee 11, Universit{\"a}t Ulm, 89069 Ulm, Germany}

\author{Martin B. Plenio}
\address{Institute of Theoretical Physics and Center for Integrated Quantum Science and Technology (IQST), Albert-Einstein-Allee 11, Universit{\"a}t Ulm, 89069 Ulm, Germany}

\begin{abstract}
Recently it was proposed to use cavity-optomechanical systems to test for quantum gravity corrections to quantum canonical commutation relations [Nat. Phys. 8, 393-397 (2012)]. 
Improving the achievable precision of such devices represents a major challenge that we address with our present work. 
More specifically, we develop sophisticated paths in phase-space of such optomechanical system to obtain significantly improved accuracy and precision under contributions from higher-order corrections to the optomechanical Hamiltonian. 
An accurate estimate of the required number of experimental runs is presented based on a rigorous error analysis that accounts for mean photon number uncertainty, which can arise from classical fluctuations or from quantum shot noise in measurement.
Furthermore, we propose a method to increase precision by using squeezed states of light. 
Finally, we demonstrate the robustness of our scheme to experimental imperfection, thereby improving the prospects of carrying out tests of quantum gravity with near-future optomechanical technology.
\end{abstract}

\maketitle

\section{Introduction}

A key impediment towards a quantum theory of gravity is the difficulty in obtaining experimental evidence for quantum gravitational effects.
Experimental tests of quantum gravity suffer from the challenge that its observable effects are exceedingly small.
Theories of quantum gravity predict that quantum gravitational effects become relevant at the Planck scale.
Probing this scale directly requires energies of the order of Planck energy $E_{p} = 1.2 \times 10^{19}$~GeV, which is 15 orders of magnitude bigger than the energy at which the Large Hadron Collider operates.
Hence, it seems unlikely that these energy scales will be achieved in the near future and we must resort to indirect methods.

One indirect approach to probing Planck-scale effect relies on observing distant astronomical events for cosmological consequences of these effects~\cite{Amelino-Camelia2013}.
For instance, quantum gravity predicts that the velocity of photons depends on their energies.
Thus, photons travelling from distant gamma ray bursts over cosmological distances will incur a detectable spread in their arrival times of photons~\cite{Amelino-Camelia1998}.
This approach, however, suffers from challenges as it includes model-dependent assumptions, for example about the evolution of the objects that emit them and on extraneous effects in the path of the photons.
This lack of control of the experimental conditions is compounded by the limitations to possible improvements to the precision of such experiments as they are intrinsically limited by the finite size of the universe.
This motivates looking for an alternative route to detecting Planck-scale effects which allow, at least in principle, for scaling of the sensitivity of the experiment with advancing technology.

One such route involves using the remarkable precision of quantum optical, optomechanical and matter-wave devices~\cite{Albrecht2014,Bawaj2015}.
Pikovski~\emph{et al.}~propose an optomechanical scheme to test for quantum gravity effects~\cite{Pikovski2012,Bosso2016}. 
Using the prediction that the canonical commutation relations suffer corrections due to quantum gravity, this scheme proposes to measure the canonical commutator of a massive object directly.
Using an optical field, the state of the mechanical resonator is taken through a loop in phase space causing the commutator of the position and momentum operator of the mechanical oscillator to be mapped to the phase of the optical field. 
The commutator is measured and the contribution from regular quantum mechanics is subtracted to estimate the quantum gravity parameter.
Although the proposal suffers from the challenge of the so-called `soccer-ball problem'~\cite{Hossenfelder2014,Quesne2010}, i.e., it is not clear whether the deformations should apply to individual particles or the centre of mass of the mechanical resonator, the proposal is promising in exploring an entirely new parameter space at the intersection of quantum mechanics and gravity.
Furthermore, technological progress and advanced experimental protocols have the potential to improve sensitivity by many orders of magnitude.

However, there are some challenges in the analysis of the original proposal that make it difficult to realise experimentally.
The contributions from the higher order corrections to the cavity Hamiltonian are much larger than the quantum gravity signal and need to be taken into account to avoid false positives. 
The precision of the estimated parameters is reduced because of uncertainty in the incident-light mean photon number, which can arise from classical fluctuations or from quantum shot noise in measurement.
We address these issues by taking higher order terms into account and suggesting different, more complicated paths in phase space so that the imprecision arising from photon number uncertainty is minimized.
We also suggest using squeezed states of light to further improve precision.

The paper is organised as follows: in Section~\ref{Sec:Background}, we introduce the notation that will be used in this work and review the optomechanical scheme suggested by Pikovski~\emph{et al.}~\cite{Pikovski2012}. 
In Section~\ref{Sec:Problems} we discuss the analysis of Pikovski~\emph{et al.} and point out that upon careful scrutiny the original claims of accuracy and precision might need to be revised.
We show that we need to account for higher order terms of the Hamiltonian to ensure accuracy in the value of the estimated quantum gravity parameters.
Furthermore, we show that unaccounted for mean photon number uncertainty reduces the precision of the inferred quantum gravity parameters.
In Section~\ref{Sec:Results} we present sophisticated paths in phase space that minimise the magnitude of the quantum mechanical contributions to the signal (so that precision can be enhanced) and calculate the resultant unitary operator to higher accuracy.
We then calculate the required experimental parameters and show that the number of required experimental runs is significantly reduced using our paths in phase space and our refined analysis.
We also show that the precision can further be improved with the use of squeezed states and we provide an analysis to this end.
As a result of these improvements we achieve several orders of magnitude enhancements in sensitivity to the signal originating from possible quantum gravitational origin.
Finally, we conclude in Section~\ref{Sec:Conclusion}.

\section{Background}
\label{Sec:Background}
In this section, we describe the existing scheme for testing quantum gravity effects via cavity optomechanics due to Pikovski \emph{et al.}~\cite{Pikovski2012}.
We first describe the quantity that is measured i.e., the canonical commutator which is deformed due to quantum gravity. 
We then introduce relevant notation and outline the features of the scheme that pertain to our analysis. 
We conclude this section with the experimental parameters suggested by Pikovski \emph{et al.}

\subsection{Modified commutation relations}
The existence of a minimum length scale as predicted by quantum gravity requires that the canonical commutation relations are modified.
Phenomenological models of quantum gravity predict that the commutation relation of the canonical position and momentum operators, $x$ and $p$ differs from $\mathrm{i}\hbar$ by a small correction term. 
Different models of quantum gravity predict different correction terms, characterised by weights that we describe in the remainder of the section and refer to as quantum gravity parameters. 
Pikovski \emph{et al.}~\cite{Pikovski2012} suggest a scheme that measures this canonical commutator directly, thereby estimating the quantum gravity parameters.

One model of quantum gravity~\cite{Kempf1995} leads to a deformed commutator of the form 
\begin{equation}
\left[ x,p \right]_{\beta_0} = \iota \hbar \left(1+  \beta_0 \left(\frac{p}{ M_{P} \, c} \right)^2 \right) 
\label{Eq:Beta0Commutator}
\end{equation}
where the strength of the correction is characterised by the constant quantum gravity parameter~$\beta_{0}$. 
The other constants in the expression are the Planck constant $M_{P}$ and the speed of light $c$.
Another model~\cite{Maggiore1993} of quantum gravity leads to the generalised version of the commutator deformation 
\begin{equation}
\left[ x,p \right]_{\mu_0} = \iota \hbar \left( 1+ 2 \mu_0 \frac{(p/c)^2 + m^2}{ M_{P}^2} \right)^{\frac{1}{2}}
\end{equation}
with correction strength given by constant $\mu_{0}$.
Notice that this deformation depends on the mass $m$ of the particle. 
In the limit $m \ll p/c \lesssim M_p$, the commutator reduces to the $\beta_0$ commutator of Equation~\eqref{Eq:Beta0Commutator}. 
So, in existing and current analyses, we consider the other limit where $p/c \ll m \lesssim M_p$ in which case the commutator reduces to 
\begin{equation}
[x,p]_{\mu_0} = \iota \hbar \left( 1+ \mu_0 \frac{m^2}{M_p^2} \right)
\label{Eq:Mu0Commutator}
\end{equation}
which is what we consider in the remainder of this paper.
Another recently proposed model~\cite{Ali2009} of quantum gravity leads to the commutator deformation 
\begin{equation}
\left[ x,p \right]_{\gamma_0} = \iota \hbar \left(1 - \gamma_0 \frac{p}{ M_{P} \, c} + \gamma_0^2 \left(\frac{p}{ M_{P} \, c} \right)^2 \right) 
\end{equation}
with quantum gravity parameter $\gamma_{0}$.
We consider the limit $\gamma_0 \ll 1$ or $p/c \ll M_p$, where the commutator reduces to 
\begin{equation}
[x,p]_{\gamma_0} = \iota \hbar \left( 1- \gamma_0 \frac{p}{M_p c} \right).
\label{Eq:Gamma0Commutator}
\end{equation}
Pikovski \emph{et al.}~propose an experimental scheme to measure the values of $\beta_{0}$, $\mu_{0}$ and $\gamma_{0}$. 
We describe the scheme in the next section.

\subsection{Experimental scheme by Pikovski~\textit{et al.}}
In this section, we detail the experimental scheme used by Pikovski \emph{et al.}. 
We later describe the challenges in their scheme and suggest modifications to overcome these challenges.

The scheme of Pikovski \emph{et al.}~relies on light interacting with a mechanical resonator via a cavity field. 
We describe the interaction Hamiltonian in Section~\ref{Sec:Hamiltonian}.
The phase acquired by the output light depends on the commutator of the resonator's position and momentum. 
In Section~\ref{Sec:MeasuringDeformations} we detail the experimental sequence to imprint the commutator on the phase of the output light. 
Measuring this phase enables testing quantum gravity corrections to the canonical commutation relations.
Finally, we describe their analysis of required experimental parameters in Section~\ref{Sec:P_Experiment}.

\subsubsection{The Hamiltonian}
\label{Sec:Hamiltonian}
The Hamiltonian that couples the light and mechanical resonator is given by
\begin{equation}
H = \hbar \omega_{m} n_{m}
+ \hbar \frac{c n_{0}}{2(L+x)} a^\dagger a,
\label{Eq:CavityHamiltonianExact}
\end{equation}
where $n_{m}$ is the number operator of the mechanical modes, $L$ is the length of the cavity at zero displacement, $x$ is the position operator describing the displacement from the mean position of the mirror, the integer $n_{0}$ depends on the frequency of the light incident at the cavity and $a$ and $a^{\dag}$ are the annihilation and creation operators of light modes.

The Hamiltonian is approximated by expanding to first order in $x$ as
\begin{equation}
H \approx \hbar \omega_{m} n_{m}
+ \hbar \omega_L a^\dagger a 
- \hbar \omega_L \frac{x}{L} a^\dagger a, 
\label{Eq:HamiltonianApprox}
\end{equation}
where
\begin{equation}
\omega_L = \frac{cn_{0}}{2L}.
\end{equation}
Rewriting the Hamiltonian in terms of the dimensionless quadratures (satisfying the commutation relations given by Equation~\eqref{Eq:betashort})
\begin{align}
X &= x \left( \frac{\hbar}{m\omega_m} \right)^{-\frac{1}{2}} \\
P &= p \left( \hbar m\omega_m \right)^{-\frac{1}{2}}
\label{Eq:Dimensionless}
\end{align}
where $m$ is the mass of the mirror and $\omega_{m}$ is the frequency of the mirror, and defining
\begin{equation}
g_0 = \omega_L \left( \frac{\hbar}{m\omega_m L^2} \right)^{\frac{1}{2}}
\end{equation}
we rewrite the Hamiltonian as 
\begin{equation}
H \approx \hbar \omega_{m} n_{m}
+ \hbar \omega_L a^\dagger a 
- \hbar g_0 X a^\dagger a.
\label{Eq:HamiltonianOrderOne}
\end{equation}
For sufficiently short pulses, the first term can be ignored. 
In an interaction picture with respect to $H_{0} = \hbar \omega_L a^\dagger a $ the evolution of the optical degree of freedom is determined by the last term.

In Section~\ref{Sec:Problems}, we show that the approximation~\eqref{Eq:HamiltonianApprox} of truncating the Hamiltonian to only the first order is not valid under the conditions that are relevant to the detection of possible quantum gravitational corrections to the canonical commutation relations.
Nonetheless, we demonstrate the implications of this assumption in the remainder of this section to describe the experimental scheme~\cite{Pikovski2012}.

\subsubsection{The scheme: measuring deformations in the commutator}
\label{Sec:MeasuringDeformations}
The Hamiltonian given by Equation~\eqref{Eq:HamiltonianOrderOne} along with pulsed optomechanics~\cite{Vanner2011} is used to take the state of the mechanical oscillator through a loop in phase space i.e., the state is acted upon by the unitary operator 
\begin{equation}
U = \e^{\iota \lambda n P}\e^{-\iota \lambda n X}\e^{-\iota \lambda n P}\e^{\iota \lambda n X}
\label{Eq:XiLoop}
\end{equation}
where $\lambda \simeq g_{0}/ \kappa$ with $\kappa$ is the optical amplitude decay rate and $n = a^\dagger a$.
This loop ensures that the phase of the measured light field of the cavity depends on the canonical commutator. 
$\lambda$ depends on the finesse $\mathcal{F}$ of the cavity as $\lambda = 4 \mathcal{F} x_{0}/ \lambda_{L}$ where $\lambda_{L}$ is the optical wavelength and $x_{0} =\left( \frac{\hbar}{m\omega_m} \right)^{-\frac{1}{2}} $.

\begin{figure}[h]\centering
 \includegraphics[height=0.3\columnwidth]{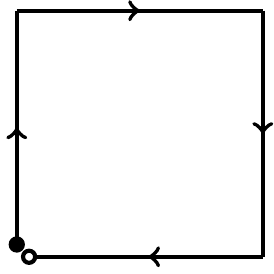}
\caption{$U$: The loop in phase space to measure quantum gravity parameters that manifest in corrections to the canonical commutation relations.}
\label{Fig:UPikovski}
\end{figure}

The four-displacement operator~\eqref{Eq:XiLoop} is calculated for the commutator of the form
\begin{equation}
[X, P] = \iota \left(1 + \beta P^2\right),
\label{Eq:betashort}
\end{equation}
which is obtained from Equation~\eqref{Eq:Beta0Commutator} using dimensionless position and momentum operators~\eqref{Eq:Dimensionless} and defining $\beta = \beta_0 \frac{\hbar \omega_m m}{M_p c}$. 
The four-displacement operator is calculated to be
\begin{equation}
U = \e^{-\iota \lambda^2 n^2}\e^{-\iota \beta \left(\lambda^2 n^2 P^2 + \lambda^3 n^3 P + \left(1/3 \right) \lambda^4 n^4 \right)}
\end{equation}
up to first order in $\beta$.

During the experiment, the mean optical field of the outgoing light is measured. 
It is given by 
\begin{equation}
\braket{a} = \mathrm{Tr}\left( a U \rho_{m}^{th} \otimes \rho_{\ell}^{\alpha} U^{\dagger} \right)
\end{equation}
for initial mechanical and optical state $\rho_{m}^{th}$ and $\rho_{\ell}^{\alpha}$ respectively.
The initial state of the mechanical oscillator is assumed to be the thermal state
\begin{equation}
\rho^{th}_m = \sum_{n_m = 0}^{\infty} \frac{\bar{n}^{n_m}}{(1+\bar{n})^{1+n_m}} \ket{n_m}\bra{n_m}
\end{equation}
where $\bar{n}$ is the mean phonon number of the oscillator.
The state of light is initially in a coherent state given by
\begin{equation}
\rho^\alpha_\ell = \ket{\alpha}\bra{\alpha} = \e^{-|\alpha|^2}\sum_{n_\ell,k_\ell}\frac{\alpha^{n_\ell}\alpha^{*k_\ell}}{\sqrt{n_\ell! k_\ell!}}\ket{n_\ell}\bra{k_\ell}
\end{equation}
where $\ket{n_\ell}$ are the Fock states.
For $N_p = |\alpha|^{2} \gg 1$ and $\lambda^{2} N_p^{3}\gg \bar{n}$, the mean optical field can be approximated by 
\begin{equation}
\braket{a_\ell} \approx \alpha \e^{-\iota\lambda^2 - N_p(1 - \e^{-i2\lambda^2})} \e^{-\iota\Theta_{\beta}}
\end{equation}
where $\Theta_{\beta}$ is given by
\begin{equation}
\Theta_{\beta} \approx \frac{4}{3}\beta N_p^3 \lambda^4 \e^{-\iota 6\lambda^2}.
\end{equation}
We see that the phase of the state of light measured has contribution from the quantum gravity corrections to the commutator.

Similar calculations can be performed with the $\gamma_{0}$~\eqref{Eq:Gamma0Commutator} and $\mu_{0}$~\eqref{Eq:Mu0Commutator} commutators, which can be rewritten as
\begin{equation}
[X, P] = \iota \left(1 - \gamma P \right) 
\quad \mathrm{for} \quad 
\gamma = \gamma_{0} \frac{ \sqrt{\hbar m \omega}}{M_p c} 
\end{equation}
and 
\begin{equation}
[X, P] = \iota \left(1 + \mu \right)
\quad \mathrm{for} \quad 
\mu = \mu_{0} \frac{m^{2}}{M_{p}^{2}}.
\end{equation}
These calculations show that 
\begin{equation}
\Theta_{\gamma} \approx - \frac{3}{2}\gamma N_p^2 \lambda^3 \e^{-\iota 4\lambda^2}
\end{equation}
and
\begin{equation}
\Theta_{\mu} \approx 2 \mu N_p \lambda^2 \e^{-\iota 2\lambda^2}.
\end{equation}
In order to estimate the contribution from quantum gravity, the total phase is measured experimentally and the quantum mechanical contribution $\lambda^2 -\iota N_p(1 - \e^{-i2\lambda^2}) \approx 2 N_{p} \lambda^{2}$ is subtracted from the total phase to get $\Theta_{\beta / \gamma / \mu}$.

The above calculations of the phases are only valid when we assume that the cavity Hamiltonian can be truncated to first order in $x$~\eqref{Eq:HamiltonianApprox}. 
Before describing the challenges of the assumption in Section~\ref{Sec:Problems}, we describe experimental requirements in next subsection.

\subsubsection{Uncertainty analysis and required experimental parameters}
\label{Sec:P_Experiment}
The precision to which the quantum gravity parameters are determined depends on the experimental parameters used and the number of times, $N_{r}$, the experiment is performed.
The values of $\mu_{0}$, $\gamma_{0}$ and $\beta_{0}$ are expected to be of order 1~\cite{Plato2016}.
To have a precision of $\delta \mu_{0} \sim 1$, $\delta \gamma_{0} \sim 1$ and $\delta \beta_{0} \sim 1$, the required number of runs of experiment is calculated. 
It is assumed that the uncertainty in the total measured phase $\Phi_{T}$ is proportional to the 
uncertainty in the quantum gravity parameters, i.e, the other terms contribute a negligible amount of uncertainty.
We will show in Section~\ref{Sec:Problems} that this assumption is not always correct.
The number of experimental runs $N_{r}$ is calculated using the relation 
\begin{equation}
\delta \braket{\Phi_{T}} = \frac{1}{ 2 \sqrt{N_{p} N_{r}} }
\label{Eq:P_Precision}
\end{equation}
and the results are listed in Table~\ref{Tab:OldParameters}.

\begin{table}[h]
\renewcommand{\arraystretch}{1.15}
\centering
\begin{tabular}{|c |c |c |c | }
\hline
Parameters & $\mu$ equation & $\gamma$ equation &  $\beta$ equation \\
\hline
$\mathcal{F}$  &  $10^{5}$      & $2 \times 10^{5}$  & $4 \times 10^{5}$  \\
$m$       &  $ 10^{-11}$~kg   & $ 10^{-9}$~kg    & $ 10^{-7}$~kg \\
$\frac{\omega_m}{2\pi}$&  $10^{5}$~Hz     & $10^{5}$~Hz     & $ 10^{5}$~Hz \\
$\lambda_L$   &  $1064$~nm      & $1064$~nm      & $532$~nm \\
$N_p$      &  $10^{8}$      & $5 \times 10^{10}$  & $ 10^{14}$ \\
$\delta \braket{\Phi}$& $10^{-4}$     & $ 10^{-8}$  & $10^{-10}$ \\
$N_r$      &  $1$         & $10^5$        & $10^6$ \\
\hline
\end{tabular}
\caption{Experimental parameters as suggested by Pikovski \emph{et. al}}
\label{Tab:OldParameters}
\end{table}

In summary, the Pikovski \emph{et al.}~scheme measures the quantum gravity parameters by using optomechanics to emit light whose phase is proportional to the quantum gravity parameters.
The calculations of the phase of the outgoing light are performed assuming that the cavity Hamiltonian is truncated to first order in the displacement of the cavity's mirror. 
The experimental parameters required to perform this experiment are calculated assuming that the uncertainty in the mean number of photons can be ignored. 
In Section~\ref{Sec:Problems}, we show that these assumptions are not valid and suggest modifications to the scheme and to the calculations in Section~\ref{Sec:Results} to overcome these challenges.

\section{Revisiting the analysis of Pikovski \emph{et al.}}
\label{Sec:Problems}
In Section~\ref{Sec:AccuracyBackground}, we show that the terms ignored in the analysis of Ref.~\cite{Pikovski2012} in the Taylor expansion~\eqref{Eq:HamiltonianApprox} of the cavity Hamiltonian to first order in $x$ are significant. 
That is, these higher order corrections contribute significantly to the total phase and hence cannot be ignored.
In Section~\ref{Sec:PrecisionBackground}, we account for non-zero uncertainty in the mean photon number and show that when it is accounted for, the precision of the estimated parameters can decrease by several orders of magnitude. 
The precise adjustment depends on how the experiment is performed.
So, in order to have the same precision in the estimated quantum gravity parameters, we would need to repeat the experiment far more often than suggested originally.

\subsection{Accuracy}
\label{Sec:AccuracyBackground}

Here we consider the higher order corrections to the cavity Hamiltonian~\eqref{Eq:CavityHamiltonianExact} and calculate the additional phase incurred by the outgoing light due to these terms.
So we instead retain the higher order terms to obtain
\begin{equation}
H = \hbar \omega_{m} n_{m}
+ \hbar \omega_L a^\dagger a 
- \hbar g_0 X a^\dagger a 
+ \hbar g_0 k X^2 a^\dagger a 
+ \dots
\label{Eq:HamiltonianExpanded}
\end{equation}
where $k = \sqrt{\frac{\hbar}{m \omega_m L^2}}$.
To take the higher order terms into account, we define $H_{X}$ and $H_{P}$ as 
\begin{align}
\begin{split}
H_{X} =& \, n \lambda_0 \left( X - k X^2 + k^{2} X^3 - \dots \right) \\
H_{P} =& \, n \lambda_0 \left( P - k P^2 + k^{2} P^3 - \dots \right) 
\label{Eq:HxHp_main}
\end{split}
\end{align}
and calculate the four-displacement operator given by
\begin{equation}
U = \e^{\iota H_{P}} 
\e^{-\iota H_{X}} 
\e^{-\iota H_{P}} 
\e^{\iota H_{X}}.
\label{Eq:4Displacement}
\end{equation}
While we focus on the nonlinearities of the form~\eqref{Eq:HamiltonianExpanded} for concreteness, our analysis can also be used for other forms of non-linearities in $X$ in the Hamiltonian.
For example, the accuracy might possibly be improved by considering corrections arising from the microscopic Hamiltonian by generalising the procedure adopted in Ref.~\cite{Law1995} to higher powers of $X$ than unity.

The effect of some specific anharmonic terms in the Hamiltonian, namely either $X^{3}$ or $X^{4}$ terms, on the phase has been studied in Ref.~\cite{Latmiral2016} but in this case, a full analysis is required for obtaining accurate estimates of the quantum gravity parameter.

To illustrate the effect of the higher order terms and for ease of calculation, we consider the Hamiltonian expanded up to third order in $X$ and $P$.
We evaluate $U$ up to sixth order terms of the Baker-Campbell-Hausdorff (BCH) formula using Mathematica code~\cite{Machnes2017}. 
Keeping only those terms that contribute to a phase larger than the minimum phase uncertainty, the operator is now given by 
\begin{align}
U =\,& \exp \left\{-\iota \left( \phi_{QG}
+ \lambda_{0}^2 n^2
- 2 k \lambda_{0}^3 n^3 
+ 4 k^2 \lambda_{0}^4 n^{4} 
\right. \right.\nonumber \\ &  \left. \left.
+ \sqrt{2} k \lambda_{0}^2 n^{2} \left( (-1+ \iota) a_{m} + (-1- \iota) a_{m}^{\dagger} \right) 
\right. \right.\nonumber \\ &  \left. \left.
+ \frac{7}{\sqrt{2}} k^{2} \lambda_{0}^3 n^{3} \left( (1- \iota ) a_{m} + (1+ \iota ) a_{m}^{\dagger} \right) 
\right)\right\}
\end{align}
where 
\begin{equation}
 \phi_{QG} = 
 \begin{cases} 
  \frac{1}{3} \beta \lambda_0^4 n^{4} & ~\beta_{0}~\text{case} \\
  - \frac{1}{2} \gamma \lambda_0^3 n^3  & ~\gamma_{0}~\text{case} \\
  \mu \lambda_0^2 n^{2} & ~\mu_{0}~\text{case}
 \end{cases}
\end{equation}
and $a_{m}$ and $a^{\dag}_{m}$ are the are the annihilation and creation operators of the modes of the mechanical resonator.

We now calculate the phase acquired by light under the action of the above unitary operator on the system.
The mean optical field is given by
\begin{equation}
\braket{a} = \mathrm{Tr} \left( U^{\dagger} a U \ket{\alpha}\bra{\alpha} \otimes \rho^{th}_{m} \right)
\label{Eq:Optical field}
\end{equation}
which is evaluated in Appendix~\ref{App:MeanField4Displacement}
to get an expression of the form
\begin{equation}
\braket{a} = \alpha' \e^{-\iota \Phi_{T}}
\end{equation}
where
\begin{equation}
\Phi_{T} = \Phi_{QG}
+ 2 \lambda_{0}^{2} N_{p}
- 6 k \lambda_{0}^{3} N_{p}^{2}
+ 16 k^{2} \lambda_{0}^{4} N_{p}^{3}.
\label{Eq:PhiTSingle}
\end{equation}
and
\begin{equation}
 \Phi_{QG} = 
 \begin{cases} 
  \frac{4}{3} \beta \lambda_0^4 N_{p}^{3} & ~\beta_{0}~\text{case} \\
  - \frac{3}{2} \gamma \lambda_0^3 N_{p}^2  & ~\gamma_{0}~\text{case} \\
  2 \mu \lambda_0^2 N_{p} & ~\mu_{0}~\text{case}.
 \end{cases}
 \label{Eq:PhiQG}
\end{equation}
The assumptions made in the calculation of the phase and their validities are discussed in Appendix~\ref{App:Convergence}.

Comparing these results with those of Pikovski \emph{et al.}, we observe that we have the extra contribution $ -6 k \lambda_{0}^{3} N_{p}^{2} + 16 k^{2} \lambda_{0}^{4} N_{p}^{3}$.
In Table~\ref{Tab:ExtraTerms}, we evaluate the magnitude of this contribution for the experimental parameters suggested by Pikovski \emph{et al.}~and compare it to the minimum uncertainty in the phase due to quantum mechanical fluctuations and the expected magnitude of the quantum gravity signal. 
We see that these extra terms are larger than both the minimum uncertainty and the quantum gravity signal and therefore cannot be ignored.
Ignoring them leads to overestimation of the quantum gravity parameters.
\begin{table*}
\renewcommand{\arraystretch}{1.25}
\begin{tabular}{|c |c |c |c |c | }
\hline
Description & Terms & $\mu_{0}$ case & $\gamma_{0}$ case &  $\beta_{0}$ case \\
\hline
Quantum gravity phase & $\Phi_{QG}$  &  $10^{-4}$      & $4 \times 10^{-9}$  & $3 \times 10^{-10}$  \\
Min.~phase uncertainty & $ \frac{1}{2\sqrt{N_{p} N_{r}}} $&  $5 \times 10^{-7}$     & $2 \times 10^{-8}$    & $ 5 \times10^{-10}$ \\
\hline
QM phase from~\cite{Pikovski2012} & $ 2 \lambda_{0}^{2} N_{p}$  &  $ 4 \times10^{2}$      & $10^{4}$   & $ 10^{6}$  \\
From higher order terms & $ - 6 k \lambda_{0}^{3} N_{p}^{2} + 16 k^{2} \lambda_{0}^{4} N_{p}^{3} $&  $0.2$     & $45$    & $ 7 \times10^{5}$ \\
\hline
\end{tabular}
\caption{Magnitude of terms using the parameters suggested by Pikovski \emph{et al.}. Note that the contribution from the higher order terms is much larger than both the signal due to quantum gravity and the minimum phase uncertainty. }
\label{Tab:ExtraTerms}
\end{table*}

In summary, higher order terms in the cavity Hamiltonian have to be considered while calculating the quantum gravity phase $\Phi_{QG}$ from the total phase $\Phi_{T}$.
This is done in Section~\ref{Sec:Results}.

\subsection{Precision}
\label{Sec:PrecisionBackground}

Pikovski \emph{et al.}~consider the uncertainty in the measurement of the total phase, $\Delta \Phi_T$, for the calculation of precision as can be seen in Equation~\eqref{Eq:P_Precision}.
However, the uncertainty in the average number of photons in each laser pulse, $\Delta N_p$, is not considered. 
Since the experiment requires very high precision, it is crucial to also account for uncertainty in the mean photon number as we show in this section.

The analysis of Pikovski \emph{et al.}~assumes that the mean photon number is known precisely before the experiment measuring the phase and that it remains unchanged during the entire run of the experiment. 
However, on the one hand, the required precision of the mean photon number will necessitate large experimental time for its measurement.
On the other hand, even if an exceedingly precise measurement of the photon number is performed at the beginning of the experiment, lasers suffer from classical intensity fluctuations and drifts due to which the mean photon number becomes increasingly uncertain over time.
Thus, the uncertainty in the mean photon number must be accounted for.

Here we consider two schemes to account for this uncertainty.
In the first scheme, the intensity is measured repeatedly before each run of the experiment, for example by impinging the laser pulses on a low-reflectivity beamsplitter and performing intensity measurement on the reflected light, and the transmitted light is discarded (other methods for measuring mean-photon number will lead to a similar analysis).
By repeatedly measuring the light intensity, the effects of classical intensity fluctuations are eliminated because the remaining pulses, which are used in the QG parameter estimation, will have photon number close to the measured preceding pulses.
However, the mean photon number precision attained in these frequent measurements is limited by quantum shot noise, which we account for below.
In the second scheme, the laser intensity is similarly measured once with very high precision in the beginning of the experiment such that the effect of the quantum noise is minimised as we explain below.
The uncertainty in photon number is now dominated by classical fluctuation in photon number.
The actual experimental method and the error model would depend strongly on the experimental considerations, for instance the time and experimental complexity required to perform each kind of measurement in the lab and the amount of classical and quantum noise present. 
We now describe the schemes in detail.

\textbf{Quantum-noise-limited scheme:}
Here we propose a scheme in which the mean photon number is estimated by measuring the photon number before each run of the phase measurement.
Thus, the quantum gravity parameter estimation is performed before the mean photon number of the laser can fluctuate significantly.
While now the classical fluctuations do not contribute to the mean photon number uncertainty, the measured mean photon number unavoidably suffers from quantum uncertainty.
Specifically, if $R$ measurements of the photon number are made, the error in the mean photon number $\Delta N_{p}$ due to quantum uncertainty is $\sqrt{N_{p}/R}$. 
For high-intensity pulses, the uncertainty from classical fluctuations is usually much larger than the quantum uncertainty even for a single ($R=1$) photon-number measurement, in which case this model is useful as it provides a lower bound on the intensity fluctuations experienced in the experiment.
In this analysis, we consider the case of $R=1$ for simplicity.

\textbf{Classical-noise-limited scheme:}
The second scheme to measure the laser intensity precisely (using feedback and a long measurement time) once before the experiment begins. 
For this single measurement performed in the beginning of the experiment, effectively $R \to \infty$ so there is no contribution from quantum noise, and we call this scheme classical-noise limited.
We then perform the quantum gravity parameter estimation assuming that the mean photon number remains unchanged for the duration of the many runs of the experiment.
In this case, the uncertainty in mean photon number arises from classical fluctuations of the form $\Delta N_{p} = \epsilon N_p$. 
The relative error from classical fluctuations in photon number for short, high-intensity pulses (as required in the experiment) is of the order of $10^{-3}$ to $10^{-2}$ after stabilising the laser intensity. 
While presenting numerical values, we consider $\epsilon = 10^{-4}$ motivated by the assumption that this stability can be achieved in near future experiments.

Here we present an analysis of the precision of the quantum gravity parameters under both these schemes.
An outline of the calculations is as follows.
The largest quantum mechanical contribution to the total phase $\Phi_{T}$ is given by
\begin{equation}
\Phi_{QM} = 2 \lambda_{0}^{2} N_{p}
- 6 k \lambda_{0}^{3} N_{p}^{2}
+ 16 k^{2} \lambda_{0}^{4} N_{p}^{3}.
\end{equation}
We express the quantum gravity parameter as a function of the total measured phase and the average number of photons by substituting Equation~\eqref{Eq:PhiQG} in 
\begin{equation}
\Phi_{QG} = \Phi_{T} -\Phi_{QM}
\end{equation}
and use standard techniques in error propagation~\cite{Ku1966} to determine the variance in the calculated parameter. 
The variance of the estimated quantum gravity parameter is expressed as a function of the variances and covariance of the measured quantities $N_{p}$ and $\Phi_{T}$. 
The calculations for the $\gamma_{0}$ model are detailed below.

We begin by rewriting the quantum gravity contribution to the phase~\eqref{Eq:PhiQG} as
\begin{equation}
\Phi_{QG} = -\gamma_0 \kappa \lambda_0^3 N_p^2 \quad \text{where} \quad \kappa := \frac{3 \sqrt{\hbar m \omega}}{2 M_p c}.
\end{equation}
Expressing $\gamma_{0}$ in terms of $\Phi_{T}$ and $N_{p}$, we get
\begin{equation}
\gamma_0 = \frac{-1}{ \kappa \lambda_0^3} \left( \frac{\Phi_T}{N_p^2}\right) +  \frac{2}{\kappa \lambda_{0} N_{p}} - \frac{6k}{\kappa} + \frac{16 k^{2} \lambda_{0} N_{p}}{\kappa}
\end{equation}
and the variance in $\gamma_{0}$ is given by~\cite{Ku1966} 
\begin{align}
\left( \Delta \gamma_0 \right)^2 =& \, 
\left( \frac{1}{ \kappa \lambda_0^3 N_p^2} \right)^2 \left(\Delta \Phi_{T} \right)^2 
+ \left( \frac{2 \Phi_T}{ \kappa \lambda_0^3 N_p^3}  
- \frac{2}{\kappa \lambda_{0} N_{p}^{2}} 
\right. \nonumber \\
& \, \left.
+  \frac{16 k^{2} \lambda_{0}}{\kappa} \right)^{2} \left( \Delta N_p \right)^2
\end{align}
for one run of the experiment.

The incident light is in a coherent state but the outgoing light is not because its state gets distorted under the action of the four-displacement operator $U$. 
The standard deviation of $\Phi_{T}$ for such a distorted state is given by (details in Appendix~\ref{App:Distortion})
\begin{equation}
\Delta \Phi_{T} \approx \sqrt{\frac{1}{4 N_{p}} + \sin^{2} \left(\lambda_{0}^{2}+ 6 k \lambda_{0}^{3} N_{p} \right)}.
\end{equation}
The value of error in photon number depends on the experimental scheme used, as described above. 
In the quantum-noise-limited scheme, the standard deviation in the the inferred photon number is given by $\Delta N_{p} = \sqrt{N_p}$ whereas In the classical-noise-limited scheme, the uncertainty in inferred photon number is given by $\Delta N_{p} = \epsilon N_p$. 
Since the phase and intensity measurements are performed on different pulses, the covariance is zero.
We also note that for the experimental parameters suggested by Pikovski~\emph{et al.}, the effect of the distortion is negligible. 
However, we present it here for the sake of completeness.

The variance in $\gamma_0$ should ideally be calculated by measuring the values and variances of the total phase and number of photons.
However, to numerically estimate the precision, we substitute the expression for $\Phi_{T}$ from Equation~\eqref{Eq:PhiTSingle} and assume that $\gamma_{0} \sim 0$. 
For the experimental parameters suggested by Pikovski \emph{et al.}~we obtain the value of the variance $\left( \Delta \gamma_0 \right)^2 $ to be $10^{14}$ ($5 \times10^{16}$) in the quantum-noise-limited (classical-noise-limited) scheme. 
Hence, in order to have $\left( \Delta \gamma_0 \right)^2 \sim 1$, we need to perform the experiment $N_{r} = 10^{14}$ ($5 \times10^{16}$) times.

The number of experimental runs as predicted by Pikovski~\emph{et al.}~is $N_{r} = 10^{5}$.
The difference arises because the first term, with the uncertainty in phase, is considered by Pikovski \emph{et al.}~in the calculation of variance (Equation~\eqref{Eq:P_Precision}) but the term accounting for uncertainty in mean number of photons is ignored. 

Similar calculations are performed for the $\beta_{0}$ and $\mu_{0}$ cases (details in Appendix~\ref{App:BetaAndMu}) and the required number of experimental runs is listed in Table~\ref{Tab:PrecisionActual}.

\begin{table}[h]
\renewcommand{\arraystretch}{1.15}
\centering
\begin{tabular}{|c |c |c |c | }
\hline
Required number of runs & $\mu_{0}$ case & $\gamma_{0}$ case &  $\beta_{0}$ case \\
\hline
Suggested in Ref.~\cite{Pikovski2012} & $1$ & $10^5$ & $10^6$ \\
Including $\Delta N_{p} = \sqrt{N_p}$ & $ 10^{5} $ & $10^{14}$ & $10^{19}$ \\
Including $\Delta N_{p} = \epsilon N_p$ & $ 10^{5} $ & $5 \times 10^{16}$ & $10^{25}$ \\
\hline
\end{tabular}
\caption{Required number of experimental runs in Ref.~\cite{Pikovski2012} versus when accounting for uncertainty in number of photons $\Delta N_{p}$ (quantum- and classical-noise-limited schemes, with $\epsilon = 10^{-4}$) for different phenomenological models.}
\label{Tab:PrecisionActual}
\end{table}

In summary, we see that the required number of experimental runs can be many orders of magnitude larger when the uncertainty in the number of photons is accounted for.
Examining the calculations of the variances, we notice that most of the contribution to the variance in the quantum gravity parameters comes from the quantum mechanical terms. 
So, reducing the quantum mechanical contribution can reduce the variance, and therefore the number of runs required to attain a set precision.
In the next section, we use different paths in phase space to reduce the quantum mechanical contribution and hence the variance.

\section{Results: Phase space paths to reduce required number of experimental runs}
\label{Sec:Results}
In Section~\ref{Sec:Problems} we showed that the higher-order cavity Hamiltonian terms need to be accounted for to ensure the accurate estimation of the quantum gravity parameters. 
We also showed that depending on the available experimental parameters and the type of measurement performed, the number of required experimental runs can be orders of magnitude larger than that estimated by Pikovski \emph{et al.} and thus increase the challenges involved in the realisation of the experiment.

In this section, we suggest a way to make the scheme experimentally feasible by reducing the required number of experimental runs. 
Specifically, we suggest paths in phase space that reduce the required number of experimental runs reduces by many orders of magnitude.
We also ensure that the calculated quantum gravity parameters are accurate by taking into account the higher order terms of the cavity Hamiltonian.
To further decrease the required number of runs, we show that we can use squeezed states as the incident light as opposed to coherent states. 

The remainder of this section is organised as follows.
We first describe the path in phase space that reduces the required number of runs.
We then calculate the phase acquired by light due to the action of the unitary operator that effects this path.
From the expression of the acquired phase, we calculate the variance in the estimated QG parameter and therefore the required number of runs for the same experimental parameters as before and show that the number of runs is many orders of magnitude smaller.
We then show that using squeezed states can further reduce the required number of runs.

For illustration, we focus on the $\gamma_{0}$ case.
The path to reduce the number of runs is composed of four rectangular loops.
Each of the loops is similar to that described by Equation~\eqref{Eq:XiLoop}, but starts at a different point on the rectangle, sometimes even outside the rectangle.
This four-loop path in phase space corresponds to the unitary operator
\begin{equation}
U_{\gamma_{0}} = U_{1} U_{2}^{\dag} U_{3}^{\dag} U_{4}
\label{Eq:GammaFourLoop}
\end{equation}
where the individual components are given by
\begin{align}
\begin{split}
U_{1} = &\,
\e^{- 2 \iota H_{X}}
\e^{- \iota H_{P}} 
\e^{ \iota H_{X}} 
\e^{ \iota H_{P}} 
\e^{ \iota H_{X}} \\
U_{2} = &\,
\e^{- \frac{7}{3} \iota H_{X}}
\e^{- \iota H_{P}} 
\e^{\iota H_{X}} 
\e^{ \iota H_{P}} 
\e^{ \frac{4}{3}\iota H_{X}} \\
U_{3} = &\,
\e^{ \frac{2}{3} \iota H_{P}} 
\e^{-\iota H_{X}}
\e^{-\iota H_{P}} 
\e^{\iota H_{X}}
\e^{ \frac{1}{3} \iota H_{P}}\\
U_{4} = &\,
\e^{\iota H_{P}} 
\e^{-\iota H_{X}}
\e^{-\iota H_{P}} 
\e^{\iota H_{X}}.
\end{split}
\end{align}
When the four loops are put together to obtain the composite loop, some parts of the path cancel and the final path is depicted in Figure~\ref{Fig:UGamma}.
The steps to arrive at such a path are detailed in Appendix~\ref{App:StepsToGetLoops}.
Depending on the coherence time of the experimental setup, we can also design paths that are made of smaller or larger number of loops by following the steps detailed in the Appendix.

Experimental realisation of a square path in phase space (Figure~\ref{Fig:UPikovski}) can be performed using a pulsed optomechanics setup described by Pikovski~\emph{et al.}~\cite{Pikovski2012}.
Specifically, the transformation of Figure~\ref{Fig:UPikovski} is implemented by alternating between phase-space translations along $X$ and $P$ axes using an optical loop to introduce time delays.
The composite rectangular paths in phase space~(Figure~\ref{Fig:UGamma}) of our proposal needs variable time delays, which can be realised by introducing an additional optical loop into the Pikovski~\emph{et al.}~setup. 
This additional loop is required to be connected to the original optical loop with fast switching, which can be implemented for instance by electro-optical modulation~\cite{Schreiber2012}.

\begin{figure}[h]\centering
 \includegraphics[height=0.7\columnwidth]{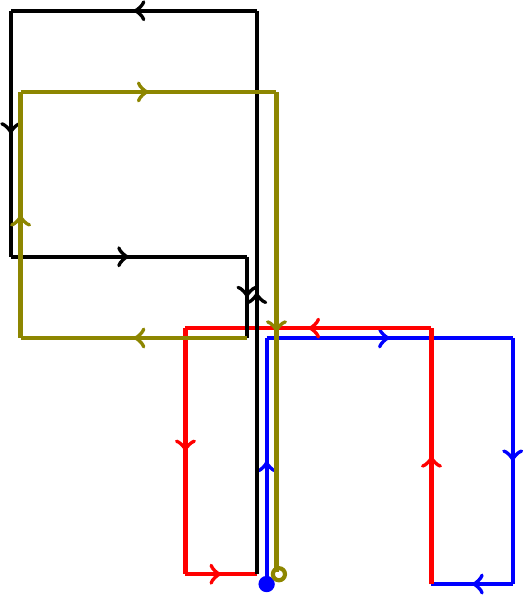}
\caption{$U_{\gamma_{0}}$: 
The final shape of the path in phase space to remove QM contribution for the $\gamma$ commutator. 
The system starts at the filled dot and ends at the unfilled dot. 
The paths are staggered for clarity, but actually overlap.}
\label{Fig:UGamma}
\end{figure}

We calculate the phase acquired by light due to the action of $U_{\gamma_{0}}$ by first expressing $U_{\gamma_{0}}$ as a single exponential by evaluating the BCH formula up to the sixth order.
The phase acquired by the outgoing light is then calculated from the resultant unitary operator by following calculations similar to those in Appendix~\ref{App:MeanField4Displacement}. 
The phase is evaluated to be
\begin{align}
\Phi_{T} =& \, \gamma \lambda_0^3 N_{p}^{2} 
- \frac{200}{3} k^3 \lambda_{0}^5 N_{p}^4
+ 144 k^4 \lambda_{0}^6 N_{p}^5
\nonumber \\
& 
+ \frac{4840}{9} k^5 \lambda_{0}^7 N_{p}^6.
\label{Eq:GammaPhiT}
\end{align}
In the calculation of the phase, several assumptions have been made.
Details about these assumptions and a discussion regarding their validity are presented in Appendix~\ref{App:Convergence}.

The parameter $\gamma_{0}$ is estimated from the total measured phase by subtracting the rest of the terms (that arise from quantum mechanics alone). 
The variance in the estimated $\gamma_{0}$ for one run of the experiment is calculated below.
If the experiment is performed $N_{r}$ number of times, the variance reduces by a factor of $N_{r}$.
We calculate the number of runs required to for the variance to be of order 1, i.e., $\left( \Delta \gamma_{0} \right)^{2} \sim 1$.

Here we calculate the uncertainty in $\gamma_0$ assuming that we know $\lambda_0$ exactly, but neither the total measured phase $\Phi_T$ nor the average number of photons in the optical state $N_p$.
The calculations and assumptions here are similar to those in Section~\ref{Sec:PrecisionBackground}.

In order to estimate the variance, we use Equation~\eqref{Eq:GammaPhiT}, to express $\gamma_{0}$ as a function of the experimentally measured quantities $\Phi_{T}$ and $N_{p}$.
\begin{align}
\gamma_0 =& \frac{1}{ \kappa \lambda_0^3} \left( \frac{\Phi_T}{N_p^2}\right) 
+ \frac{200 \lambda_0^{2} k^{3}}{3 \kappa } N_p^{2}
- \frac{144 \lambda_0^{3} k^{4}}{ \kappa } N_p^{3}
\nonumber \\
& \, 
- \frac{4840 \lambda_0^{4} k^{5}}{9 \kappa } N_p^{4}.
\end{align}
where 
\begin{equation}
\kappa := \frac{\sqrt{\hbar m \omega}}{M_p c}.
\end{equation}
Using standard techniques in error propagation~\cite{Ku1966}, we determine the uncertainty in $\gamma_{0}$ to be
\begin{align}
& \left( \Delta \gamma_0 \right)^2 =  \left( \frac{1}{ \kappa \lambda_0^3 N_p^2} \right)^2 \left(\Delta \Phi_{T} \right)^2 
+ \left( - \frac{2 \Phi_T}{ \kappa \lambda_0^3 N_p^3} 
\right.
\nonumber \\
& \quad \left.
+ \frac{400 \lambda_0^{2} k^{3}}{3 \kappa} N_p
- \frac{432 \lambda_0^{3} k^{4}}{ \kappa } N_p^{2}
\right)^2 \left( \Delta N_p \right)^2
\label{Eq:GammaUncertainty}
\end{align}
for one run of the experiment.
The uncertainty in photon number is 
\begin{equation}
\Delta N_{p} = \sqrt{N_{p}} \quad \mathrm{or} \quad \Delta N_{p} = \epsilon N_{p}
\end{equation}
depending on the experimental scheme used.
The state of light after the action of the unitary operator is no longer coherent but distorted.
Hence the standard deviation of $\Phi_{T}$ for such a distorted state is given by (details in Appendix~\ref{App:Distortion})
\begin{equation}
\Delta \Phi_{T} \approx \sqrt{\frac{1}{4 N_{p}} + 
\sin^{2} \left( 
360 k^{4} \lambda_{0}^{6} N_{p}^{4}
-\frac{400}{3} k^{3} \lambda_{0}^{5} N_{p}^{3}  \right)}.
\end{equation}
We estimate the value of the variance for experimental parameters suggested by Pikovski~\emph{et al.}~and obtain $\left( \Delta \gamma_0 \right)^2 = 6 \times 10^{5}$ in both schemes.
The value is the same in both schemes because we have now successfully eliminated contribution from $\Delta N_p$ terms for these experimental parameters and all the contribution is from $\Delta \Phi_T$ terms.
The number of experimental runs required to have $\left( \Delta \gamma_0 \right)^2 =1$ is $N_{r} = 6 \times 10^{5}$, as opposed to $10^{14}$ or $5 \times10^{16}$ runs required if we perform only the single loop.
We also note that for the given experimental parameters, the effect of the distortion of the state is negligible but is presented here for completeness.

\begin{figure}[h]
\centering
\includegraphics[width=0.95\columnwidth]{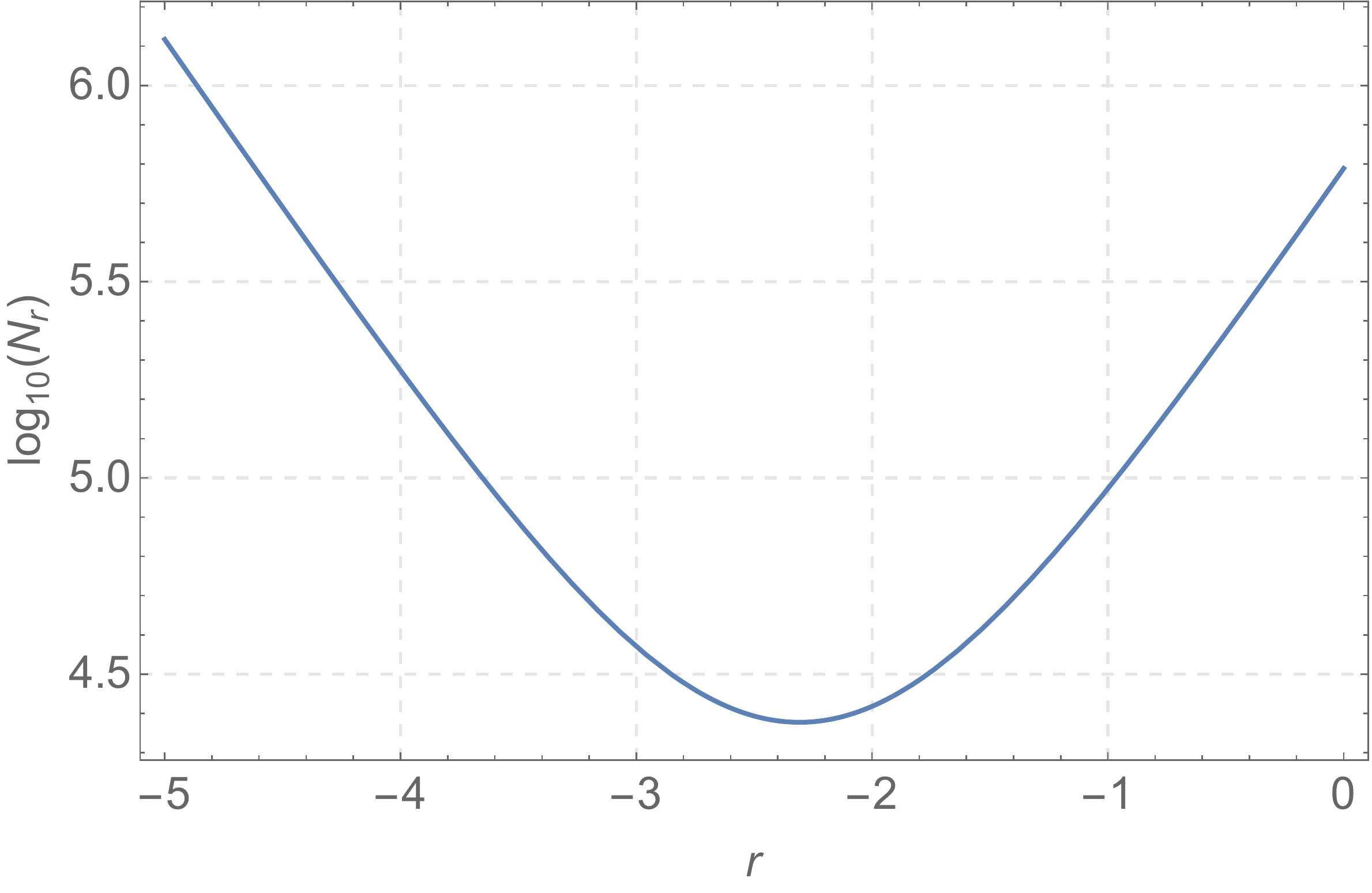}
\caption{ $\log_{10}(N_{r})$ as a function of the squeezing parameter $r$ for fixed experimental parameters}
\label{Fig:GammaNvsr}
\end{figure}

The precision in the estimated QG parameter can be increased further by using squeezed states of light. 
We see from Equation~\eqref{Eq:GammaUncertainty} that the variance in $\gamma_{0}$ depends both on $\Delta \Phi_{T}$ and $\Delta N_{p}$. 
Plugging in the experimental parameters suggested by Pikovski~\emph{et al.}, we see that in the $\gamma_{0}$ case, the contribution from the $\Delta \Phi_{T}$ term is the largest. 
Therefore, we can perform the experiment using light squeezed in $\Phi_{T}$ so that $\Delta \Phi_{T}$ is reduced at the expense of $\Delta N_{p}$ thereby improving precision. 
This is quantitatively illustrated for the $\gamma_{0}$ case below where we see an order of magnitude improvement in precision.

Assuming that the effects of distortion are negligible, the variances for ideal squeezed states with large number of photons are approximately (See Ref.~\cite{Kim1989} and Appendix~\ref{App:StatisticsSqueezedStates} for calculations)
\begin{equation}
\left(\Delta \Phi_{T} \right)^2 \approx \frac{\e^{2r}}{4 N_p}, 
\enskip \left( \Delta N_p \right)^2 \approx N_{p} \e^{-2r}, 
\end{equation}
where $r$ is the squeezing parameter. 
Keeping all parameters the same as those suggested by Pikovski~\emph{et al.}~and using squeezed light with squeezing parameter $r = -2.3$ gives $N_{r} = 2 \times 10^{4}$, which is an order of magnitude improvement over using coherent light.
The dependence of the number of experimental runs (Equation~\eqref{Eq:GammaUncertainty}) required is plotted as a function of the squeezing parameter in Figure~\ref{Fig:GammaNvsr}.
Similar calculations for the $\beta_{0}$ and $\mu_{0}$ cases are presented in Appendix~\ref{App:BetaAndMu}.
In these cases, the contribution to $\Delta \beta_{0}$ and $\Delta \mu_{0}$ is dominated by the $\Delta N_{p}$ contribution, as detailed in the appendix. 
Hence, it is useful to use light squeezed in photon number.

We also verify the robustness of our scheme to experimental imperfection. 
Specifically, we consider area-preserving fluctuations in the phase-space loops.
As detailed in~Appendix~\ref{App:Fluctuations}, we quantify the deviation in the acquired phase under these fluctuations.
We provide sufficient conditions on the magnitude of these fluctuations such that the effects of these fluctuations can be ignored.
Furthermore, we consider the effect of imperfections in the prepared thermal state of the mechanical resonator. 
In~Appendix~\ref{App:Thermal}, we detail the phase deviation due to small non-zero off-diagonal terms in the density matrix of the prepared thermal state corresponding to unintended coherences in the system.
We show that for sufficiently small off-diagonal elements, the measured phase does not differ significantly from the expected phase.

In summary, we can increase the sensitivity of the experiment to possible quantum gravity effects by using sophisticated paths in phase space and using squeezed light. 
These changes significantly improve the prospects for realising tests of quantum gravity experimentally with near-future quantum technology.

\section{Conclusion}
\label{Sec:Conclusion}
In conclusion, we address the challenge of improving the accuracy and precision of cavity-optomechanical tests of quantum gravity.
On one hand, unaccounted for mean photon number uncertainty and quantum mechanical contributions to the phase lead to low precision, while on the other, higher order terms of cavity Hamiltonian lead to low accuracy via unaccounted phase.
We account for the higher-order terms and develop sophisticated paths in phase-space to obtain experimentally feasible accuracy and precision, and we suggest the use of squeezed light to further improve precision.
Considering the quantum-noise-limited scheme, where the intensity is measured throughout the experiment, our proposed phase-space paths and rigorous analysis reduces the number of experimental runs from $10^{14}$ to $10^{5}$ for the case of the $\gamma_{0}$ model for the same experimental parameters as in the original proposal.
Similarly, considering the classical-noise-limited scheme, where the intensity is measured precisely at the beginning of the experiment, the required number of experimental runs reduces from $10^{16}$ to $10^{5}$. 
These values are calculated under the assumption that the relative error in laser intensity is $10^{-4}$ over a few hours, which might be attainable in the near future for the short, high intensity pulses that are required in this experiment.
For the $\beta_{0}$ and $\mu_{0}$ models, our suggested paths are similar to the original path. 
However, our refined analysis can help us choose better experimental parameters. 
With these parameters, the required number of experimental runs decreases by three and five orders of magnitude for the $\beta_{0}$ and $\mu_{0}$ cases respectively.

By improving the accuracy and the required number of runs, and by accounting for experimental imperfections, our work opens the way for tests of quantum gravity with near-future optomechanical technology.

\section*{Acknowledgements}
The authors are grateful to S.~Machnes for making available the QLib Mathematica package which was used in calculation of BCH-expansions in this work. 
The authors would also like to thank F.~Armata, T.~Tufarelli, V.~Tkachuk and M.~Vanner for useful comments on the draft.
The authors acknowledge high-performance computing support by the state of Baden-W\"urttemberg through bwHPC.
This work is supported by the EU projects DIADEMS and the ERC Synergy grant BioQ.

\appendix

\section{Calculating $\braket{a}$}
\label{App:MeanField4Displacement}
In this section, we detail the calculations in calculating the phase acquired by the outgoing light from the pulse sequence that acts on the system.

Starting from the given pulse sequence, we use Mathematica code~\cite{Machnes2017} to express the product of exponentials as a single exponential using the BCH formula.
That is, we simplify the product of exponentials 
\begin{equation}
U = \e^{\iota H_{P}} 
\e^{-\iota H_{X}} 
\e^{-\iota H_{P}} 
\e^{\iota H_{X}}
\label{Eq:ProductU}
\end{equation}
to a single exponential of the form
\begin{align}
U =& \, \exp \left\{ -\iota \phi_{QG} -\iota w(n) + \left( x^{*}n^{2} + y^{*}n^{3} \right) a^{\dagger}_{m} 
\right. \nonumber\\
& \, \left.
- \left( x n^{2} + y n^{3}\right) a_{m}+ \dots \right\}.
\label{Eq:Uexact}
\end{align} 
In order to carry out this simplification, we first need to truncate the Hamiltonian to a finite order in $k$. 
$H_{X}$ and $H_{P}$ are given by
\begin{align}
\begin{split}
H_{X} =& \, \lambda_0 n_{\ell} \left( X_{m} - k X_{m}^2 + k^{2} X_{m}^3 - \dots \right) \\
H_{P} =& \, \lambda_0 n_{\ell} \left( P_{m} - k P_{m}^2 + k^{2} P_{m}^3 - \dots \right).
\label{Eq:App_HXHP}
\end{split}
\end{align}
and we truncate it to a finite order in $k$.
The simplification of $U$ to a single exponential is then carried out using the BCH formula evaluated to a finite order in BCH order.
The simplified expression so obtained~\eqref{Eq:Uexact} still has a large number of terms in the exponential. 
Calculation of the mean field $\braket{a}$ from Equation~\eqref{Eq:Uexact} is difficult. 
So we only keep those terms in $U$ that contribute to a significant phase and neglect the phase contribution from the rest of the terms.
These assumptions and approximations entering this step are discussed in detail in~Appendix~\ref{App:Convergence}.

We now calculate the mean field $\braket{a}$ of the light to estimate the phase of light.
In these calculations, we ignore $\phi_{QG}$ and calculate the phase from only the quantum mechanical terms.
The unitary operator after truncation of terms is
\begin{equation}
U = \e^{-i w(n) + \left( x^{*}n^{2} + y^{*}n^{3} \right) a^{\dagger}_{m} - \left( x n^{2} + y n^{3}\right) a_{m}}
\label{Eq:UTrunc}
\end{equation} 
where 
\begin{align}
w(n) &= \lambda_{0}^2 n^2 - 2 k \lambda_{0}^3 n^3 + 4 k^2 \lambda_{0}^4 n^{4}, \\
x &= \left( -1- i \right) \sqrt{2} k \lambda_{0}^2, \\ 
y &= \left( 1 + i \right) \frac{7}{\sqrt{2}} k^2 \lambda_{0}^3. 
\end{align}

The quantity that is measured is the expectation value of the annihilation operator on light states which is given by
\begin{equation}
\braket{a} = \mathrm{Tr} \left( U^{\dagger} a U \ket{\alpha}\bra{\alpha} \otimes \rho^{th}_{m} \right).
\end{equation}
In the remainder of this section, we calculate the above quantity. 

We begin by rewriting $U$ as
\begin{equation}
U = \e^{-i w(n)} \e^{\left( x^{*}a^{\dagger}_{m} - x a_{m} \right) n^{2} + \left( y^{*}a^{\dagger}_{m} - y a_{m} \right) n^{3}} 
\label{Eq:Split1}
\end{equation}
and using the Zassenhaus formula~\cite{Casas2012}, simplifying to 
\begin{equation}
U = 
\e^{\left( x^{*}a^{\dagger}_{m} - x a_{m} \right) n^{2}} 
\e^{\left( y^{*}a^{\dagger}_{m} - y a_{m} \right) n^{3}} 
\e^{-\frac{1}{2} n^{5} \left( x^{*}y - x y^{*}\right)}
\e^{ -i w(n) }.
\label{Eq:Split2}
\end{equation}
Therefore $U^{\dagger} a U$ is given by 
\begin{align}
U^{\dagger} a U =& \,
\e^{i w(n)}
\e^{\frac{1}{2} n^{5} \left( x^{*}y - x y^{*}\right)}
\e^{-\left( y^{*}a^{\dagger}_{m} - y a_{m} \right) n^{3}} 
\nonumber\\
& 
\times
\e^{-\left( x^{*}a^{\dagger}_{m} - x a_{m} \right) n^{2}} 
a 
\e^{\left( x^{*}a^{\dagger}_{m} - x a_{m} \right) n^{2}} 
\nonumber\\
& 
\times
\e^{\left( y^{*}a^{\dagger}_{m} - y a_{m} \right) n^{3}} 
\e^{-\frac{1}{2} n^{5} \left( x^{*}y - x y^{*}\right)}
\e^{-i w(n)}.
\end{align}
First evaluate $\e^{-\left( x^{*}a^{\dagger}_{m} - x a_{m} \right) n^{2}} a \e^{\left( x^{*}a^{\dagger}_{m} - x a_{m} \right) n^{2}} $ from the expression for $U^{\dagger} a U$ using the BCH formula 
\begin{align}
\e^{X}Y \e^{-X} =& Y+\left[X,Y\right]+\frac{1}{2!}[X,[X,Y]]
\nonumber\\
& \,
+\frac{1}{3!}[X,[X,[X,Y]]]+\dots.
\end{align}
to obtain
\begin{align}
& \e^{-\left( x^{*}a^{\dagger}_{m} - x a_{m} \right) n^{2}}  a \e^{\left( x^{*}a^{\dagger}_{m} - x a_{m} \right) n^{2}} = 
\nonumber \\
& \qquad
a 
-\left( x^{*}a^{\dagger}_{m} - x a_{m} \right) \left[ n^{2},a \right] 
\nonumber \\
& \qquad
+\frac{1}{2!} \left( x^{*}a^{\dagger}_{m} - x a_{m} \right)^{2} \left[ n^{2}, \left[ n^{2},a \right] \right] + \dots. .
\end{align}
Observing that 
\begin{equation}
\left[ n^{2},a \right] = - \left( 2n +1 \right) a
\end{equation}
and simplifying, we get
\begin{equation}
\e^{-\left( x^{*}a^{\dagger}_{m} - x a_{m} \right) n^{2}} a \e^{\left( x^{*}a^{\dagger}_{m} - x a_{m} \right) n^{2}} = 
\e^{\left( x^{*}a^{\dagger}_{m} - x a_{m} \right) \left( 2n +1 \right) } a. 
\end{equation}
Now $U^{\dagger} a U$ reads as follows:
\begin{align}
U^{\dagger} a U =& \,
\e^{i w(n)}
\e^{\frac{1}{2} n^{5} \left( x^{*}y - x y^{*}\right)}
\e^{-\left( y^{*}a^{\dagger}_{m} - y a_{m} \right) n^{3}} 
\nonumber\\
& 
\times
\e^{\left( x^{*}a^{\dagger}_{m} - x a_{m} \right) \left( 2n +1 \right) }
a 
\e^{\left( y^{*}a^{\dagger}_{m} - y a_{m} \right) n^{3}} 
\nonumber\\
& 
\times
\e^{-\frac{1}{2} n^{5} \left( x^{*}y - x y^{*}\right)}
\e^{-i w(n)}.
\end{align}

To perform similar calculations for the $y$ terms, we should first interchange the terms $\e^{-\left( y^{*}a^{\dagger}_{m} - y a_{m} \right) n^{3}}$ and $\e^{\left( x^{*}a^{\dagger}_{m} - x a_{m} \right) \left( 2n +1 \right) }$. Using the Zassenhaus formula again, we have
\begin{align}
& \e^{-\left( y^{*}a^{\dagger}_{m} - y a_{m} \right) n^{3}} 
\e^{\left( x^{*}a^{\dagger}_{m} - x a_{m} \right) \left( 2n +1 \right) }
=
\e^{\left( x^{*}a^{\dagger}_{m} - x a_{m} \right) \left( 2n +1 \right) }
\nonumber\\
& \qquad
\times
\e^{-\left( y^{*}a^{\dagger}_{m} - y a_{m} \right) n^{3}} 
\e^{\left( x^{*}y-xy^{*}\right) \left( 2n^{4}+ n^{3} \right)}.
\end{align}
Now we evaluate $\e^{-\left( y^{*}a^{\dagger}_{m} - y a_{m} \right) n^{3}} a \e^{\left( y^{*}a^{\dagger}_{m} - y a_{m} \right) n^{3}}$ similarly as in the $x$ case using the BCH formula to get
\begin{align}
& \e^{-\left( y^{*}a^{\dagger}_{m} - y a_{m} \right) n^{3}} a \e^{\left( y^{*}a^{\dagger}_{m} - y a_{m} \right) n^{3}} = 
\nonumber \\
& \qquad
a -\left( y^{*}a^{\dagger}_{m} - y a_{m} \right) \left[ n^{3},a \right] 
\nonumber \\
& \qquad
+\frac{1}{2!} \left( y^{*}a^{\dagger}_{m} - y a_{m} \right)^{2} \left[ n^{3}, \left[ n^{3},a \right] \right] + \dots. .
\end{align}
Using the formula
\begin{equation}
\left[ n^{3},a \right] = - \left( 3n^{2} + 3n +1 \right) a
\end{equation}
and simplifying, we find
\begin{align}
\e^{-\left( y^{*}a^{\dagger}_{m} - y a_{m} \right) n^{3}} & a \e^{\left( y^{*}a^{\dagger}_{m} - y a_{m} \right) n^{3}} =
\nonumber\\
& \e^{\left( y^{*}a^{\dagger}_{m} - y a_{m} \right) \left( 3n^{2} + 3n +1 \right) } a. 
\end{align}
Now $U^{\dagger} a U$ is given by
\begin{align}
U^{\dagger} a U =& \,
\e^{\left( x^{*}y-xy^{*}\right) \left( 2n^{4} + n^{3} \right)}
\e^{i w(n)}
\e^{\frac{1}{2} n^{5} \left( x^{*}y - x y^{*}\right)}
\nonumber\\
& 
\times
\e^{\left( x^{*}a^{\dagger}_{m} - x a_{m} \right) \left( 2n +1 \right) }
\e^{\left( y^{*}a^{\dagger}_{m} - y a_{m} \right) \left( 3n^{2} + 3n +1 \right) }  
\nonumber\\
& 
\times
a
\e^{-\frac{1}{2} n^{5} \left( x^{*}y - x y^{*}\right)}
\e^{-i w(n)}.
\end{align}

Using similar techniques, we evaluate $\e^{\frac{1}{2} n^{5} \left( x^{*}y - x y^{*}\right)} a \e^{-\frac{1}{2} n^{5} \left( x^{*}y - x y^{*}\right)}$ and $\e^{i w(n)} a \e^{-i w(n)}$. 
Observing that 
\begin{equation}
\left[ n^{5},a \right] = - \left( 5n^{4} +10 n^{3} +10 n^{2} + 5n + 1 \right) a,
\end{equation}
we simplify
\begin{align}
& \e^{\frac{1}{2} n^{5} \left( x^{*}y - x y^{*}\right)} a \e^{-\frac{1}{2} n^{5} \left( x^{*}y - x y^{*}\right)} =
\nonumber\\
& \qquad
 \e^{-\frac{1}{2}\left( x^{*}y - x y^{*}\right) \left( 5n^{4} +10 n^{3} +10 n^{2} + 5 n + 1 \right) } a.
\end{align}
and
substituting for $w(n)$ and observing that
\begin{equation}
\left[ n^{4},a \right] = - \left( 4 n^{3} +6 n^{2} + 4n + 1 \right) a,
\end{equation}
we obtain 
\begin{align}
\e^{i w(n)} a \e^{-i w(n)} = & \,
\e^{i 2 k \lambda_{0}^{3} \left( 3n^{2} + 3n +1 \right)} 
\e^{- i \lambda_{0}^{2} \left( 2n +1 \right)} 
\nonumber \\
& 
\times
\e^{- i 4 k^{2} \lambda_{0}^{4} \left( 4n^{3} + 6n^{2} + 4n +1 \right)}
a.
\end{align}
Now $U^{\dagger} a U$ is given by
\begin{align}
U^{\dagger} a U =& \,
\e^{\left( x^{*}y-xy^{*}\right) \left( 2n^{4} + n^{3} \right)}
\e^{\left( x^{*}a^{\dagger}_{m} - x a_{m} \right) \left( 2n +1 \right) }
\nonumber \\
&
\times
\e^{\left( y^{*}a^{\dagger}_{m} - y a_{m} \right) \left( 3n^{2} + 3n +1 \right) } 
\e^{i 2 k \lambda_{0}^{3} \left( 3n^{2} + 3n +1 \right)} 
\nonumber \\
&
\times
\e^{-\frac{1}{2}\left( x^{*}y - x y^{*}\right) \left( 5n^{4} +10 n^{3} +10 n^{2} + 5 n + 1 \right) }
\nonumber \\
&
\times
\e^{- i \lambda_{0}^{2} \left( 2n +1 \right)} 
\e^{- i 4 k^{2} \lambda_{0}^{4} \left( 4n^{3} + 6n^{2} + 4n +1 \right)}
a.
\end{align}
which can be re-written as
\begin{align}
U^{\dagger} a U
= & \,
\e^{-\frac{1}{2}\left( x^{*}y - x y^{*}\right)\left( n^{4} + 8 n^{3} +10 n^{2} + 5n + 1 \right)}
\nonumber \\
&
\times
\e^{i 2 k \lambda_{0}^{3} \left( 3n^{2} + 3n +1 \right)} 
\e^{- i \lambda_{0}^{2} \left( 2n +1 \right)} 
\nonumber \\
&
\times
\e^{- i 4 k^{2} \lambda_{0}^{4} \left( 4n^{3} + 6n^{2} + 4n +1 \right)}
\e^{\left( x^{*}a^{\dagger}_{m} - x a_{m} \right) \left( 2n +1 \right) }
\nonumber \\
&
\times
\e^{\left( y^{*}a^{\dagger}_{m} - y a_{m} \right) \left( 3n^{2} + 3n +1 \right) } 
a.
\end{align}

We now calculate the quantity of interest - the expectation value of the annihilation operator on light states. 
Note that $ a \ket{\alpha} = \alpha \ket{\alpha} $. 
By definition, $\braket{a}$ is given by
\begin{align}
\braket{a} = & \,\mathrm{Tr} \left( 
\e^{-\frac{1}{2}\left( x^{*}y - x y^{*}\right)\left( n^{4} + 8 n^{3} +10 n^{2} + 5n + 1 \right)}
\e^{- i \lambda_{0}^{2} \left( 2n +1 \right)} 
\right.
\nonumber\\
& 
\times
\e^{i 2 k \lambda_{0}^{3} \left( 3n^{2} + 3n +1 \right)} 
\e^{- i 4 k^{2} \lambda_{0}^{4} \left( 4n^{3} + 6n^{2} + 4n +1 \right)}
\nonumber\\
& 
\times
\e^{\left( x^{*}a^{\dagger}_{m} - x a_{m} \right) \left( 2n +1 \right) }
\e^{\left( y^{*}a^{\dagger}_{m} - y a_{m} \right) \left( 3n^{2} + 3n +1 \right) } 
\nonumber\\
& \, \left.
\times
\alpha
\ket{\alpha}\bra{\alpha} \otimes \rho^{th}_{m} 
\right).
\end{align}
Writing the trace explicitly, we have
\begin{align}
\braket{a} = & \, \sum_{m=0}^{\infty} 
\frac{\bar{n}^{m}}{(1+\bar{n})^{1+m}}
\bra{\alpha, m}
\alpha
\e^{- i 4 k^{2} \lambda_{0}^{4} \left( 4n^{3} + 6n^{2} + 4n +1 \right)}
\nonumber\\
& 
\times
\e^{-\frac{1}{2}\left( x^{*}y - x y^{*}\right)\left( n^{4} + 8 n^{3} +10 n^{2} + 5n + 1 \right)}
\e^{- i \lambda_{0}^{2} \left( 2n +1 \right)} 
\nonumber\\
& 
\times
\e^{i 2 k \lambda_{0}^{3} \left( 3n^{2} + 3n +1 \right)} 
\e^{\left( x^{*}a^{\dagger}_{m} - x a_{m} \right) \left( 2n +1 \right) }
\nonumber\\
& 
\times
\e^{\left( y^{*}a^{\dagger}_{m} - y a_{m} \right) \left( 3n^{2} + 3n +1 \right) } 
\ket{\alpha, m}.
\end{align}
We simplify the above expression in the remainder of this section. 
Inserting identities $\sum_{k=0}^{\infty} \ket{k} \bra{k}$ and $\sum_{n=0}^{\infty} \ket{n} \bra{n}$ in the Hilbert space of the light field, 
\begin{align}
\braket{a} = & \, \sum_{m=0}^{\infty} 
\sum_{k,n=0}^{\infty} 
\frac{\bar{n}^{m}}{(1+\bar{n})^{1+m}}
\nonumber\\
& 
\times
\bra{\alpha, m}
\alpha
\e^{-\frac{1}{2}\left( x^{*}y - x y^{*}\right)\left( n^{4} + 8 n^{3} +10 n^{2} + 5n + 1 \right)}
\nonumber\\
& 
\times
\e^{- i 4 k^{2} \lambda_{0}^{4} \left( 4n^{3} + 6n^{2} + 4n +1 \right)}
\e^{i 2 k \lambda_{0}^{3} \left( 3n^{2} + 3n +1 \right)} 
\nonumber\\
& 
\times
\e^{- i \lambda_{0}^{2} \left( 2n +1 \right)} 
\ket{k,m}
\bra{k,m}
\e^{\left( x^{*}a^{\dagger}_{m} - x a_{m} \right) \left( 2n +1 \right) }
\nonumber\\
& 
\times
\e^{\left( y^{*}a^{\dagger}_{m} - y a_{m} \right) \left( 3n^{2} + 3n +1 \right) } 
\ket{n,m}\braket{n,m|\alpha, m}
\end{align}
and using the relationship $\braket{k|n} = \delta_{n,k}$, we have
\begin{align}
\braket{a} = & \, \sum_{m=0}^{\infty} 
\sum_{n=0}^{\infty} 
\frac{\bar{n}^{m}}{(1+\bar{n})^{1+m}} \braket{\alpha | n}
\e^{i 2 k \lambda_{0}^{3} \left( 3n^{2} + 3n +1 \right)} 
\nonumber\\
& 
\times
\alpha
\e^{-\frac{1}{2}\left( x^{*}y - x y^{*}\right)\left( n^{4} + 8 n^{3} +10 n^{2} + 5n + 1 \right)}
\e^{- i \lambda_{0}^{2} \left( 2n +1 \right)} 
\nonumber\\
& 
\times
\e^{- i 4 k^{2} \lambda_{0}^{4} \left( 4n^{3} + 6n^{2} + 4n +1 \right)}
\bra{m}
\e^{\left( x^{*}a^{\dagger}_{m} - x a_{m} \right) \left( 2n +1 \right) }
\nonumber\\
& 
\times
\e^{\left( y^{*}a^{\dagger}_{m} - y a_{m} \right) \left( 3n^{2} + 3n +1 \right) } 
\ket{m}\braket{n|\alpha}.
\end{align}

Now we evaluate $$\bra{m} \e^{\left( x^{*}a^{\dagger}_{m} - x a_{m} \right) \left( 2n +1 \right) }\e^{\left( y^{*}a^{\dagger}_{m} - y a_{m} \right) \left( 4n^{3} + 6n^{2} + 4n+ 1\right) } \ket{m}.$$
For ease of notation, we define the variables
\begin{align}
\upsilon &= y \left( 3n^{2} + 3n +1 \right) \\
\chi &= x \left( 2n +1 \right).
\end{align}
We denote the displaced Fock state $\e^{\left( \upsilon^{*} a^{\dagger}_{m} - \upsilon a_{m} \right)} \ket{m}$ as $\ket{\upsilon^{*}, m}$. 
By definition
\begin{align}
\bra{m} \e^{\left( \chi^{*}a^{\dagger}_{m} - \chi a_{m} \right)}\e^{\left( \upsilon^{*} a^{\dagger}_{m} - \upsilon a_{m} \right)} \ket{m}
 =& \, \braket{-\chi^{*} , m | \upsilon^{*}, m}.
\end{align}
Using the formula for the overlap of two displaced Fock states from Ref.~\cite{Wunsche1991}, we have
\begin{align}
& \braket{-\chi^{*} , m | \upsilon^{*}, m} = 
\nonumber\\
& \quad
\braket{-\chi^{*} | \upsilon^{*}} m! \sum_{j=0}^{m} \frac{ \left( \upsilon^{*} + \chi^{*}\right)^{m-j} \left( -\chi - \upsilon \right)^{m-j} }{j! \left( m-j \right)! \left( m-j \right)! }
\end{align}
where
\begin{equation}
 \braket{-\chi^{*} | \upsilon^{*}} = \exp \left\{ -\chi\upsilon^{*} -\frac{1}{2}\left( |\chi|^{2} + |\upsilon|^{2} \right) \right\}.
\end{equation}
We now sum over the mechanical modes in the expression for $\braket{a}$. The sum is given by
\begin{align}
& \sum_{m=0}^{\infty} \frac{\bar{n}^{m}}{(1+\bar{n})^{1+m}} \braket{-\chi^{*} , m | \upsilon^{*}, m} = 
\braket{-\chi^{*} | \upsilon^{*}} 
\nonumber\\
& \quad 
\times
\sum_{m=0}^{\infty} \sum_{j=0}^{m} 
\frac{\bar{n}^{m}}{(1+\bar{n})^{m+1}}
m! (-1)^{m-j} \frac{\left| \chi + \upsilon \right|^{2(m-j)} }{j! \left[\left( m-j \right)! \right]^{2}}
\end{align}
To evaluate the above expression, replace $m-j$ with $k$. This gives us
\begin{align}
& \sum_{m=0}^{\infty} \frac{\bar{n}^{m}}{(1+\bar{n})^{1+m}} \braket{-\chi^{*} , m | \upsilon^{*}, m} \nonumber 
\\
=& \, 
\braket{-\chi^{*} | \upsilon^{*}}
\sum_{k=0}^{m} \sum_{m=k}^{\infty} 
\frac{\bar{n}^{m}}{(1+\bar{n})^{m+1}}
m! (-1)^{k} \frac{\left| \chi + \upsilon \right|^{2k} }{\left( m-k \right)! \left( k! \right)^{2}} 
\nonumber 
\\
=& \, 
\braket{-\chi^{*} | \upsilon^{*}} 
\sum_{k=0}^{m} 
 (-1)^{k}\frac{\left| \chi + \upsilon \right|^{2k} }{k!}
\sum_{m=k}^{\infty} 
\binom{m}{k}
\frac{\bar{n}^{m}}{(1+\bar{n})^{m+1}}
\nonumber 
\\
=& \, 
\braket{-\chi^{*} | \upsilon^{*}} \sum_{k=0}^{m} 
 (-1)^{k} \frac{\left| \chi + \upsilon \right|^{2k} \bar{n}^{k} }{k!}
\nonumber 
\\
=& \, 
\braket{-\chi^{*} | \upsilon^{*}} 
\e^{-\left| \chi + \upsilon \right|^{2} \bar{n}}
\nonumber 
\\
=& \, 
\e^{ -\chi\upsilon^{*} -\frac{1}{2}\left( |\chi|^{2} + |\upsilon|^{2} \right) }
\e^{-\left| \chi + \upsilon \right|^{2} \bar{n}}.
\end{align}
Rewriting the expression back in terms of the original variables $x, y$ and $n$, we get
\begin{align}
\e^{ -\chi\upsilon^{*} -\frac{1}{2}\left( |\chi|^{2} + |\upsilon|^{2} \right) } & \e^{-\left| \chi + \upsilon \right|^{2} \bar{n}} 
=
\e^{ -xy^{*} \left( 2n +1 \right) \left( 3n^{2} + 3n+ 1\right)} 
\nonumber\\
& 
\times
\e^{-\frac{1}{2}\left( |x|^{2}\left( 2n +1 \right)^{2} + |y|^{2}\left( 3n^{2} + 3n+ 1\right)^{2} \right) }
\nonumber\\
& 
\times
\e^{-\left| x\left( 2n +1 \right) + y\left( 3n^{2} + 3n+ 1\right) \right|^{2} \bar{n}} .
\end{align}
Also note that the other terms that are in the expression for $\braket{a}$ are given by
\begin{equation}
\braket{\alpha|n} \braket{n|\alpha} = \e^{-|\alpha|^{2}} \frac{|\alpha|^{2n}}{n!}.
\end{equation}

Plugging these expressions back into the expression for $\braket{a}$, we have
\begin{align}
\braket{a} = & \, \sum_{n=0}^{\infty} 
\alpha
\e^{-|\alpha|^{2}} \frac{|\alpha|^{2n}}{n!}
\e^{-\frac{1}{2}\left( x^{*}y - x y^{*}\right)\left( n^{4} + 8 n^{3} +10 n^{2} + 5n + 1 \right)}
\nonumber\\
& 
\times
\e^{- i \lambda_{0}^{2} \left( 2n +1 \right)} 
\e^{- i 4 k^{2} \lambda_{0}^{4} \left( 4n^{3} + 6n^{2} + 4n +1 \right)}
\nonumber\\
& 
\times
\e^{i 2 k \lambda_{0}^{3} \left( 3n^{2} + 3n +1 \right)} 
\e^{ -xy^{*} \left( 2n +1 \right) \left( 3n^{2} + 3n+ 1\right)} 
\nonumber\\
& 
\times
\e^{-\frac{1}{2}\left( |x|^{2}\left( 2n +1 \right)^{2} + |y|^{2}\left( 3n^{2} + 3n+ 1\right)^{2} \right) }
\nonumber\\
& 
\times
\e^{-\left| x\left( 2n +1 \right) + y\left( 3n^{2} + 3n+ 1\right) \right|^{2} \bar{n}} .
\label{Eq:aExactSum}
\end{align}

The expression can be approximated using the saddle-point approximation (to leading order in $N_{p}$) to be
\begin{align}
\braket{a} = & \,
\alpha
\e^{-\frac{1}{2}\left( 4 |x|^{2} N_{p}^{2} + 9 |y|^{2} N_{p}^{4} \right)}
\nonumber\\
& 
\times
\e^{-\frac{1}{2}\left( x^{*}y - x y^{*}\right) N_{p}^{4} 
- i 2 \lambda_{0}^{2} N_{p}
-i 16 k^{2} \lambda_{0}^{4} N_{p}^{3}
+ i 6 k \lambda_{0}^{3} N_{p}^{2}
}
\nonumber\\
& 
\times
\e^{\left( 4 |x|^{2} N_{p}^{2} + 9 |y|^{2} N_{p}^{4} + 6 \left( xy^{*} + x^{*}y \right) N_{p}^{3} \right) \bar{n}} .
\label{Eq:aApprox}
\end{align}
The saddle point approximation may not be valid for all cases, for instance when the neglected terms are much larger than the quantum gravity signal.
In such cases, the sum of Equation~\eqref{Eq:aExactSum} should be evaluated numerically.

If $\braket{a}$ is given by 
\begin{equation}
\braket{a} = \alpha' \e^{-i \Phi_{QM}}, 
\end{equation}
the new amplitude is 
\begin{align}
\alpha' =& \alpha
\e^{-\frac{1}{2}\left( 4 |x|^{2} N_{p}^{2} + 9 |y|^{2} N_{p}^{4} \right)}
\nonumber\\
& 
\times
\e^{\left( 4 |x|^{2} N_{p}^{2} + 9 |y|^{2} N_{p}^{4} + 6 \left( xy^{*} + x^{*}y \right) N_{p}^{3} \right) \bar{n}} 
\end{align}
and the new phase is
\begin{equation}
\Phi_{QM} = \frac{1}{2 \iota}\left( x^{*}y - x y^{*}\right) N_{p}^{4} 
+ 2 \lambda_{0}^{2} N_{p}
+ 16 k^{2} \lambda_{0}^{4} N_{p}^{3}
- 6 k \lambda_{0}^{3} N_{p}^{2}
\end{equation}
which on substituting with $x$ and $y$ gives
\begin{equation}
\Phi_{QM} = 2 \lambda_{0}^{2} N_{p}
- 6 k \lambda_{0}^{3} N_{p}^{2}
+ 16 k^{2} \lambda_{0}^{4} N_{p}^{3}.
\end{equation}

Similar calculations hold for the calculation of phase from the four-loop paths.
The unitary operator $U$ is different, but the method and approximations are the same.

For example the $\gamma_{0}$ case, the unitary operator is approximated to
\begin{align}
U_{\gamma_{0}} =& \, \exp \left\{ - \iota \left( 
\frac{1}{3} \gamma \lambda_{0}^3 n^3
-\frac{40}{3} k^3 \lambda_{0}^5 n^5
+24 k^4 \lambda_{0}^6 n^6
\right. \right.
\nonumber \\
&
\left. \left. 
+ \frac{\sqrt{2}}{3} k^2 \lambda_{0}^3 n^3 \left(\left(-1-\iota \right) a-\left(1-\iota \right) a^{\dag}\right)
\right) \right\}.
\end{align}
This unitary operator implies that the measured field of light is given by the expression
\begin{align}
\braket{a} = & \, \sum_{n=0}^{\infty} 
\alpha
\e^{-|\alpha|^{2}} \frac{|\alpha|^{2n}}{n!}
\e^{- \frac{\iota}{3} \gamma \lambda_{0}^3 \left( 3n^{2} + 3n + 1\right)}
\nonumber\\
& 
\times
\e^{\iota \frac{40}{3} k^{3} \lambda_{0}^{5} \left( 5n^{4} + 10n^{3} + 10n^{2} + 5n + 1 \right)} 
\nonumber\\
& 
\times
\e^{- \iota 24 k^{4} \lambda_{0}^{6} \left( 6n^{5} + 15n^{4} + 20n^{3} + 15n^{2} + 6n + 1 \right)}
\nonumber\\
& 
\times 
\e^{- \frac{4}{9} k^{4} \lambda_{0}^{6} \left( 3n^{2} + 3n+ 1\right)^{2} \left(\bar{n} + \frac{1}{2} \right)}
\label{Eq:aGamma}
\end{align}
which can be evaluated numerically if higher accuracy is required.

Similarly, the unitaries for the $\beta_{0}$ and $\mu_{0}$ case are given by
\begin{align}
U_{\beta_{0}} =& \, 
\exp
\left\{ - \iota \left( 
\frac{1}{3} \beta \lambda_{0}^4 n^4
+ \lambda_{0}^2 n^2
+\frac{35}{6} k^4 \lambda_{0}^6 n^6
\right. \right.
\nonumber
\\
&
\left. 
\left.
+\sqrt{2} k \lambda_{0}^2 n^2 \left((-1+ \iota ) a-(1+\iota) a^{\dag}\right)
\right. \right.
\nonumber
\\
&
\left. 
\left.
+ \frac{1}{\sqrt{2}} k^2 \lambda_{0}^3 n^3 \left( (1+\iota) a + (1-\iota) a^{\dag}\right)
\right. \right.
\nonumber
\\
&
\left. 
\left.
+ 2 \sqrt{2} k^3 \lambda_{0}^4 n^4 \left((1- \iota) a+(1+ \iota) a^{\dag}\right)
\right. \right.
\nonumber
\\
&
\left. 
\left.
+ \frac{15}{\sqrt{2}} k^4 \lambda_{0}^5 n^5 \left(-(1+ \iota) a - (1- \iota) a^{\dag} \right)
\right) 
\right\}
\end{align}
and
\begin{align}
U_{\mu_{0}} =& \, 
\exp
\left\{ - \iota \left( 
\mu \lambda_{0}^2 n^2
+ \lambda_{0}^2 n^2
\right. \right.
\nonumber
\\
&
\left. 
\left.
+\sqrt{2} k \lambda_{0}^2 n^2 \left((-1+ \iota ) a-(1+\iota) a^{\dag}\right)
\right) 
\right\}
\end{align}
and the corresponding expression for mean field $\braket{a}$ can be obtained with similar calculations.

\section{Arriving at the improved loops in phase space}
\label{App:StepsToGetLoops}

Sophisticated paths in phase space are used to reduce the quantum mechanical contribution and therefore, the required number of experimental runs.
The steps to arrive at the sophisticated path are outlined here.

\begin{figure}[h]\centering
 \includegraphics[height=0.3\columnwidth]{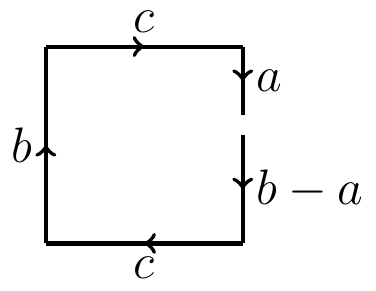}
\caption{Shape of the path in phase space corresponding to $U_{X}$}
\label{Fig:UX}
\end{figure}

\begin{figure}[h]\centering
 \includegraphics[height=0.33\columnwidth]{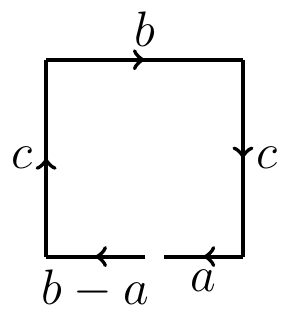}
\caption{Shape of the path in phase space corresponding to $U_{P}$}
\label{Fig:UP}
\end{figure}

First, we consider unitary operators which describe arbitrary rectangular pulse sequences.
Such unitary operators are given by $U_{X}$ and $U_{P}$, where changing the values of $a$, $b$ and $c$ changes the dimensions of the loop and also determines the starting point.
The operators are given by
\begin{equation}
U_{X} = \e^{-\iota a H_{X}} 
\e^{-\iota c H_{P}} 
\e^{\iota b H_{X}}
\e^{\iota c H_{P}} 
\e^{-\iota (b - a) H_{X}} 
\end{equation}
and
\begin{equation}
U_{P} = \e^{\iota a H_{P}} 
\e^{-\iota c H_{X}} 
\e^{- \iota b H_{P}}
\e^{\iota c H_{X}} 
\e^{\iota (b - a) H_{P}}
\end{equation}
and are represented as loops in phase space in Figures~\ref{Fig:UX} and~\ref{Fig:UP}.

We then fix the number of loops that we want the sophisticated path to be made of.
More paths can reduce the required number of runs, but they also increase the required coherence time. Also, calculating the final phase of light can be more computationally intense with a larger number of loops.
So, depending on the coherence time, the number of loops can be chosen. 
In this case, we choose four loops, two like $U_{X}$ and two like $U_{P}$.

Once the loops are chosen, we express the final unitary operator as a single exponential of a sum of operators instead of a product of exponentials using the BCH formula.
This is done using Mathematica package~\cite{Machnes2017}. 
The final simplified unitary operator is expressed as a function of the parameters $\{ a_{i}, b_{i}, c_{i}\}; i = 1,2,3,4$ for the four loops.

We order the resulting terms in order of descending magnitude of how much these terms contribute to the final phase of light.
In this ordering, we assume that the ordering is the same if we directly substitute the experimental parameters in the operators (i.e., replacing the operator $n$ with the average number of photons $N_{p}$).

Once the ordering is done, we choose values of the parameters $\{ a_{i}, b_{i}, c_{i}\}$ such that the coefficients of the largest $m$ quantum mechanical terms are zero, while the coefficient of the quantum gravity term is nonzero. 
We choose the largest $m$ possible such that the solutions $\{ a_{i}, b_{i}, c_{i}\}$ exist.
This is how we determine a path in phase space that can minimize the quantum mechanical contribution while keeping the quantum gravity contribution non-zero.


\section{Number and phase statistics of squeezed states}
\label{App:StatisticsSqueezedStates}

\begin{figure}[h]
\centering
 \includegraphics[height=0.8\columnwidth]{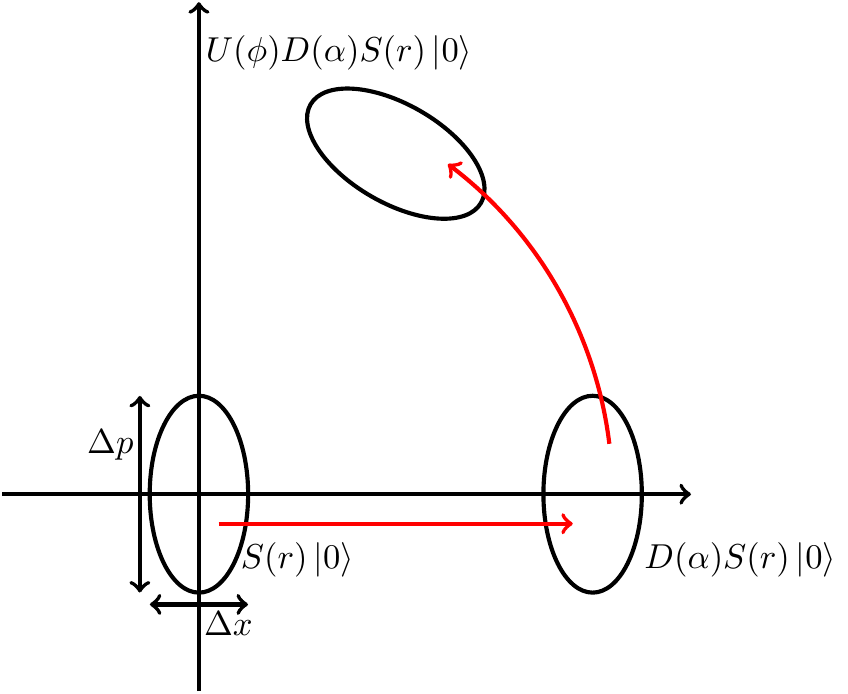}
\caption{A rotated displaced squeezed state for real displacement vector $\alpha$}
\label{Fig:SqueezedState}
\end{figure}

In this section, we recap some relevant results about number and phase properties of displaced squeezed states.
First, we introduce some notation and an assumption regarding the state of light used.
To understand how using squeezed states affects the uncertainty in the signal, we consider ideal squeezed states, which are defined as squeezed vacuum states which are displaced in phase space. 
The state is given by
\begin{equation}
\ket{\alpha, r} = D(\alpha) S(r) \ket{0}
\end{equation}
where $S(r)$ is the squeezing operator with squeezing parameter $r$
\begin{equation}
S(r) = \exp \left( \frac{r a_{m}^2 - r a_{m}^{\dagger 2}}{2}\right)
\end{equation}
and $D(\alpha)$ is the displacement operator displacement vector $\alpha$
\begin{equation}
D(\alpha) = \exp \left( \alpha a_{m}^{\dagger} - \alpha^{*} a_{m} \right)
\end{equation}
where $a^{\dagger}_{m}$ and $a_{m}$ are the creation and annihilation operators respectively.

In calculating the uncertainties in the phases due to a squeezed state, we assume that the final state to be measured can be described by an ideal squeezed state with squeezing parameter $r$ and displacement $\alpha$. The final state that is measured can be described by an ideal squeezed state if the unitary operator only rotates the state and does not distort it as illustrated in Figure~\ref{Fig:SqueezedState}.

For squeezed states, $\Delta \Phi_{T}$ is given by
\begin{equation}
\left(\Delta \Phi_{T} \right)^2 = \frac{\e^{2r}}{4 \left( N_p - \sinh^2{r} \right)}
\label{Eq:DeltaPhi}
\end{equation}
and $\Delta N_p$ given by \cite{Kim1989}
\begin{equation}
\left( \Delta N_p \right)^2 = \frac{1}{2} \sinh^2{2r} + \left(N_p - \sinh^2{r}\right)\e^{-2r}.
\label{Eq:DeltaNpPhi0}
\end{equation}

For the large values of $N_{p}$ that are used in this experiment, the uncertainties in phase and number of photons can be approximated to 
\begin{equation}
\left(\Delta \Phi_{T} \right)^2 = \frac{\e^{2r}}{4 N_p}
\end{equation}
and 
\begin{equation}
\left( \Delta N_p \right)^2 = N_{p} \e^{-2r}. 
\end{equation}

In the next subsections, we calculate the average photon number $N_p = \braket{\hat{n}}$, the uncertainty in the number of photons $\Delta N_p = \sqrt{\braket{(\Delta \hat{n})^2}}$ and the uncertainty in the total phase $\Delta \Phi_{T}$ in terms of the squeezing parameter and displacement vector. 
In order to calculate these quantities, we use the equations
\begin{align}
D^{\dagger}(\alpha) a D(\alpha) &= a + \alpha \label{Eq:Da}\\
D^{\dagger}(\alpha) a^{\dagger} D(\alpha) &= a^{\dagger} + \alpha^*, \label{Eq:DaDag}
\end{align}
and
\begin{align}
S^{\dagger}(r) a S(r) &= a \cosh{r} - a^{\dagger} \sinh{r} \label{Eq:Sa}\\
S^{\dagger}(r) a^{\dagger} S(r) &= a^{\dagger} \cosh{r} - a\sinh{r} \label{Eq:SaDag}.
\end{align}

\subsection{Calculation of average photon number in a displaced squeezed state}
The average photon number is given by
\begin{align}
N_p =& \braket{\hat{n}} \\
=& \braket{\alpha, r | a^{\dagger}a | \alpha, r}\\
=& \braket{0 | S^{\dagger} D^{\dagger} a^{\dagger}a D S | 0}.
\end{align}
We now evaluate $D^{\dagger} a^{\dagger}a D$ using Equations~\eqref{Eq:Da} and \eqref{Eq:DaDag} to obtain
\begin{align}
D^{\dagger} a^{\dagger}a D =& D^{\dagger} a^{\dagger} D D^{\dagger}a D\\
=& \left( a^{\dagger} + \alpha^* \right) \left( a + \alpha \right)\\
=& a^{\dagger}a + \alpha a^{\dagger} + \alpha^* a + |\alpha|^2. \label{Eq:DaaD}
\end{align}
Using Equations~\eqref{Eq:Sa} and \eqref{Eq:SaDag}, we see that
\begin{align}
S^{\dagger} D^{\dagger} a^{\dagger} & a D S
\nonumber \\  
=& \, S^{\dagger} a^{\dagger}a S+ \alpha S^{\dagger} a^{\dagger} S + \alpha^* S^{\dagger} a S + |\alpha|^2 S^{\dagger} S\\
=& \left( a^{\dagger} \cosh{r} - a\sinh{r} \right) \left( a \cosh{r} - a^{\dagger} \sinh{r}\right) 
 \nonumber \\ & 
+ \alpha \left( a^{\dagger} \cosh{r} - a\sinh{r} \right)
 \nonumber \\ & 
 + \alpha^* \left( a \cosh{r} - a^{\dagger} \sinh{r}\right) + |\alpha|^2.
\end{align}
The surviving terms in $ \braket{0 | S^{\dagger} D^{\dagger} a^{\dagger}a D S | 0}$ are
\begin{equation}
\braket{0 | S^{\dagger} D^{\dagger} a^{\dagger}a D S | 0} = \braket{0 | a a^{\dagger} | 0} \sinh^2{r} + |\alpha|^2.
\end{equation}
This gives
\begin{equation}
N_p = |\alpha|^2 +\sinh^2{r}. 
\label{Eq:Np_squeezed}
\end{equation}

\subsection{Calculation of spread in photon number in a displaced squeezed state}
The uncertainty in the photon number is defined as
\begin{align}
\Delta N_p =& \sqrt{\braket{(\Delta \hat{n})^2}}\\
=& \sqrt{ \braket{\hat{n}^2} - \braket{\hat{n}}^2 }.
\end{align}
We begin by evaluating $\braket{\hat{n}^2}$. Writing it explicitly, we have
\begin{align}
\braket{\hat{n}^2} =& \braket{\alpha, r | a^{\dagger}a a^{\dagger}a | \alpha, r}\\
=& \braket{0 | S^{\dagger} D^{\dagger} a^{\dagger}a a^{\dagger}a D S | 0}.
\end{align}
We evaluate $D^{\dagger} a^{\dagger}a a^{\dagger}a D$ using Equation~\eqref{Eq:DaaD} to obtain
\begin{align}
D^{\dagger} a^{\dagger}a a^{\dagger}a D = & \left( a^{\dagger}a + \alpha a^{\dagger} + \alpha^{*} a + |\alpha|^2 \right)^2. 
\end{align}
Only terms with even number of operators in the above expression contribute to the calculation of $\braket{\hat{n}^2}$. Keeping only such contributing terms, we get
\begin{align}
\braket{\hat{n}^2} =& 
\alpha^2 \braket{0 | S^{\dagger}a^{\dagger 2} S | 0} 
+\alpha^{* 2} \braket{0 | S^{\dagger}a^2 S | 0} 
+ | \alpha|^2
+ |\alpha|^4
\nonumber \\
&
+ \braket{0 | S^{\dagger}a^{\dagger}a a^{\dagger}a S | 0} 
+ 4 |\alpha|^2 \braket{0 | S^{\dagger}a^{\dagger}a S | 0}. 
\end{align}
Using Equations~\eqref{Eq:Sa} and \eqref{Eq:SaDag} and simplifying, we find
\begin{align}
\braket{\hat{n}^2} =& \, 2 \sinh^2{r}\cosh^2{r} + \sinh^4{r}
- \alpha^{* 2} \sinh{r}\cosh{r}
\nonumber \\
&
- \alpha^2 \sinh{r}\cosh{r} 
+ 4 |\alpha|^2 \sinh^2{r}
\nonumber \\
&
+ |\alpha|^2
+ |\alpha|^4.
\end{align}

To calculate $\braket{\hat{n}}^2$, recall that from Equation~\eqref{Eq:Np_squeezed} we have
\begin{equation}
\braket{\hat{n}} = |\alpha|^2 +\sinh^2{r}. 
\end{equation}
We now calculate the variance in the photon number to be
\begin{align}
\braket{(\Delta \hat{n})^2} =& \braket{\hat{n}^2} - \braket{\hat{n}}^2 \\
=& \, 2 \sinh^2{r}\cosh^2{r} + \sinh^4{r}
- \alpha^{* 2} \sinh{r}\cosh{r}
\nonumber \\
&
- \alpha^2 \sinh{r}\cosh{r} 
+ 4 |\alpha|^2 \sinh^2{r}
+ |\alpha|^2
+ |\alpha|^4 
\nonumber \\
&
- \left( |\alpha|^4 +\sinh^4{r} + 2 |\alpha|^2 \sinh^2{r} \right) \\
=& \, 2 \sinh^2{r}\cosh^2{r} 
+ |\alpha|^2
- \alpha^{* 2} \sinh{r}\cosh{r}
\nonumber \\
&
- \alpha^2 \sinh{r}\cosh{r} 
+ 2 |\alpha|^2 \sinh^2{r}.
\end{align}
Writing $\alpha := |\alpha|\e^{i\phi}$, we rewrite the above expression as
\begin{align}
\braket{(\Delta \hat{n})^2}  = & \frac{1}{2} \sinh^2{2r} 
+ |\alpha|^2 \left( 1 + 2\sinh^2{r} 
\right. \nonumber \\
&  \left.
- 2 \sinh{r}\cosh{r} \cos{2\phi} \right)
\end{align}
which can be rewritten as
\begin{equation}
\braket{(\Delta \hat{n})^2} = \frac{1}{2} \sinh^2{2r} + |\alpha|^2 \left( \e^{2r}\sin^2{\phi} + \e^{-2r}\cos^2{\phi} \right).
\end{equation}
Therefore,
\begin{equation}
\Delta N_p = \sqrt{ \frac{1}{2} \sinh^2{2r} + |\alpha|^2 \left( \e^{2r}\sin^2{\phi} + \e^{-2r}\cos^2{\phi} \right) }
\label{Eq:DelNp}
\end{equation}
which matches the expression of \cite{Kim1989}.
We consider real displacements.
Thus, we set $\phi = 0$ and obtain 
\begin{equation}
\Delta N_p = \sqrt{ \frac{1}{2} \sinh^2{2r} + |\alpha|^2\e^{-2r}},
\label{Eq:DelNpPhi0}
\end{equation}
which, after substituting~\eqref{Eq:Np_squeezed}, is the same as Equation~\eqref{Eq:DeltaNpPhi0}.

\subsection{Calculation of uncertainty in measuring total phase for squeezed light}
The uncertainty in measuring the total phase $\Phi_T$ is the spread in the coherent state in the tangential direction (along $\Phi$) divided by the length of the vector, $|\alpha|$. Since a global phase and displacement does not alter the squeezing, we can instead consider a squeezed vacuum state to measure the spread in the $P$ quadrature. 
The $P$ quadrature is given by 
\begin{equation}
P = \left( a - a^{\dagger}\right)/2i
\end{equation}
and the spread in the state is given by
\begin{equation}
\Delta P = \sqrt{ \braket{P^2} - \braket{P}^2 }.
\end{equation}

The mean of the $P$ quadrature is zero, as can be seen from Equations~\eqref{Eq:Sa} and \eqref{Eq:SaDag}. Explicitly,
\begin{align}
\braket{P} =& \left( \braket{ 0| S^{\dagger}a S |0} - \braket{ 0| S^{\dagger}a^{\dagger} S |0} \right)/2i\\
=& \,0.
\end{align}
We now calculate $\braket{P^2}$ as
\begin{align}
\braket{P^2} =& \, \frac{1}{4} \braket{ 0| S^{\dagger} \left( 1+ 2 a^{\dagger} a - a^2 -a^{\dagger 2} \right) S |0} \\
=& \, \frac{1}{4} \left( 1+ 2\sinh^2{r} + 2\sinh{r}\cosh{r} \right)\\
=& \, \frac{1}{4} \e^{2r}.
\end{align}
Therefore,
\begin{equation}
\Delta P = \frac{1}{2} \e^{r}.
\end{equation}
Putting it all together we get,
\begin{equation}
\Delta \Phi_T = \frac{1}{2 |\alpha|} \e^{r}
\end{equation}
which, on substituting from Equation~\eqref{Eq:Np_squeezed} gives 
\begin{equation}
\Delta \Phi_T = \frac{\e^{r}}{2 \sqrt{N_p - \sinh^2{r}} }
\label{Eq:DelPhiT}
\end{equation}
which remains positive by virtue of Equation~\eqref{Eq:Np_squeezed}.
This is the uncertainty in $\Phi_T$ when only one measurement is made.

\section{State distortion}
\label{App:Distortion}
Here we consider the case when the initial state of light is a coherent state $\ket{\alpha}$.
Due to nonlinearities in the photon number $n$ in the Hamiltonian, the unitary operator acting on the initial state of light distorts the state instead of simply rotating the state. 
This leads to the variance in the phase, $\Delta \Phi$, to change. 
In this section, we calculate the value of $\Delta \Phi$ for this distorted state.

The outline of the calculations is as follows. 
The initial state of light is in a coherent state $\ket{\alpha}$ for real $\alpha$ with average photon number $N_{p} = |\alpha|^{2}$. 
The unitary operator that acts on the state during the experiment is given by $\e^{\iota f(n)}$ and we assume $f(n)$ to be a polynomial in $n$. 
If $f(n)$ is not linear in $n$, the coherent state is distorted in addition to being rotated. 
To calculate the distortion, we bring the state back to the $X$ axis and calculate the spread in $P$, $\Delta P$. 
We assume that $\Delta \Phi \approx \frac{\Delta P}{\sqrt{N_{p}}}$.

The calculations are detailed here.
To bring the state back to the $X$ axis, we calculate the phase $\Phi(N_{p})$ of the state $\e^{\iota f(n)} \ket{\alpha}$ and rotate the state back by angle $\Phi(N_{p})$. 
The state on the $X$ axis is given by
\begin{equation}
\ket{\xi} = \e^{\iota \left\{ f(n) - \Phi(N_{p}) n \right\} } \ket{\alpha} =: U \ket{\alpha}.
\end{equation}
Here, the function $\Phi(n)$ is calculated from $f(n)$ by replacing $n^{m}$ by $(n+1)^{m} - n^{m}$ for all non-zero $m$.

The variance $\left( \Delta P \right)^{2}$ is calculated by
\begin{equation}
\left( \Delta P \right)^{2} = \braket{P^{2}}_{\xi} - \braket{P}^{2}_{\xi}.
\end{equation}
We first calculate $\braket{P}_{\xi}$ which in terms of $a$ and $a^{\dag}$ is 
\begin{equation}
\braket{P}_{\xi} = \frac{1}{2 \iota} \braket{a - a^{\dag}}_{\xi}.
\end{equation}
Writing in terms of the initial coherent state, we have 
\begin{equation}
\braket{P}_{\xi} = \frac{1}{2 \iota} \left( \braket{ \alpha | U^{\dag} a U | \alpha} - \braket{ \alpha | U^{\dag} a^{\dag} U | \alpha} \right). 
\end{equation}
We can show that
\begin{equation}
U^{\dag} a U = \e^{\iota \left\{ \Phi(n) - \Phi(N_{p}) \right\} } a
\end{equation}
and making the saddle point approximation, we can approximate
\begin{equation}
\braket{ \alpha | \e^{\iota \Phi(n)} | \alpha} \approx \e^{\iota \Phi(N_{p})}. 
\end{equation}
Therefore,
\begin{equation}
\braket{P}_{\xi} = \frac{1}{2 \iota} \left( \alpha - \alpha^{*} \right)
\end{equation}
which for real $\alpha$ gives
\begin{equation}
\braket{P}_{\xi} = 0.
\end{equation}

We now calculate $\braket{P^{2}}_{\xi}$. 
\begin{align}
\braket{P^{2}}_{\xi} =& \, -\frac{1}{4} \braket{a^{2} + a^{\dag 2} - 2 a^{\dag} a -1 }_{\xi}
\nonumber \\
=& \, -\frac{1}{4} \left( \braket{ \alpha | U^{\dag} a^{2} U | \alpha} + \braket{ \alpha | U^{\dag} a^{\dag 2} U | \alpha} 
\right. \nonumber \\ & \quad \left.
-2 \braket{ \alpha | U^{\dag} a^{\dag}a U | \alpha} -1 \right)
\end{align}
$U$ commutes with $a^{\dag} a = n$, so $U^{\dag} a^{\dag}a U = a^{\dag}a$ and therefore $\braket{ \alpha | U^{\dag} a^{\dag}a U | \alpha} = N_{p}$. 
We now evaluate $U^{\dag} a^{2} U$. We observe
\begin{align}
U^{\dag} a^{2} U =& \, \left( U^{\dag} a U \right)^{2}
\nonumber \\
=& \, \e^{\iota \left\{ \Phi(n) - \Phi(N_{p}) \right\} } a \, \e^{\iota \left\{ \Phi(n) - \Phi(N_{p}) \right\} } a
\nonumber \\
=& \, \e^{2 \iota \left\{ \Phi(n) - \Phi(N_{p}) \right\} } \e^{\iota \Theta(n)} a^{2} 
\end{align}
where the function $\Theta(n)$ is calculated from $\Phi(n)$ by replacing $n^{m}$ by $(n+1)^{m} - n^{m}$ for all non-zero $m$.
Therefore
\begin{align}
\braket{P^{2}}_{\xi} =& \, -\frac{1}{4} \left( N_{p} \e^{\iota \Theta(N_{p})} + N_{p} \e^{-\iota \Theta(N_{p})} - 2 N_{p} -1 \right)
\nonumber \\
=& \, \frac{1}{4} \left( 1 + 4 N_{p} \sin^{2} \frac{\Theta(N_{p})}{2} \right).
\end{align}
This leads to
\begin{equation}
(\Delta \Phi)^{2} = \frac{1}{4 N_{p}} + \sin^{2} \frac{\Theta(N_{p})}{2}.
\end{equation}


\section{Area preserving fluctuations}
\label{App:Fluctuations}
We consider the phase acquired by the light after performing the paths in phase space, assuming that the paths are subject to imperfections, i.e., assuming that state of the mechanical resonator undergoes area-preserving fluctuations in phase space.
We provide sufficient conditions for these deformations to have a negligible effect on the estimation of the quantum gravity signal.

\begin{figure*}
\centering
\hspace{-2ex}
\begin{subfigure}{0.4\textwidth}
 \includegraphics[width=0.6\columnwidth]{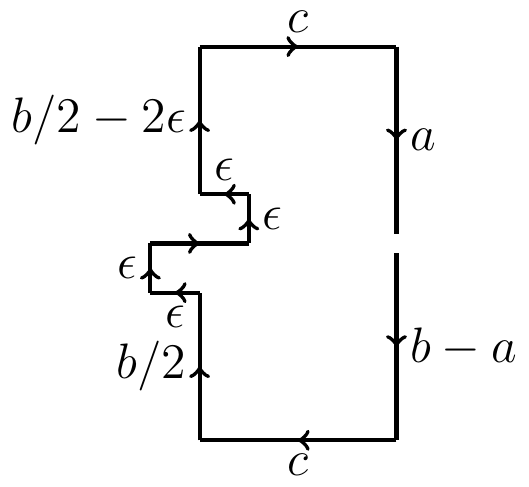}
 \caption{}
\label{Fig:X_fluctX}
\end{subfigure}
\begin{subfigure}{0.4\textwidth}
 \includegraphics[width=0.5\columnwidth]{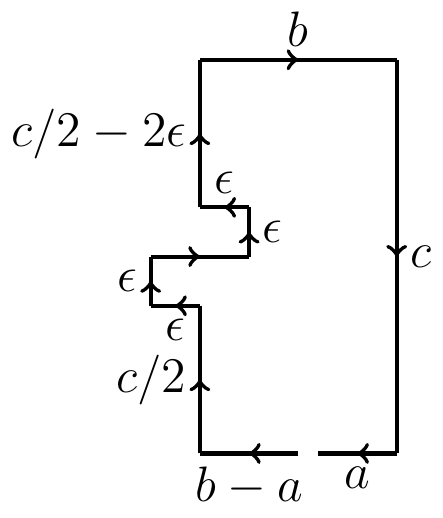}
 \caption{}
 \label{Fig:P_fluctX}
\end{subfigure}

\hspace{-2ex}
\begin{subfigure}{0.4\textwidth}
 \includegraphics[width=0.7\columnwidth]{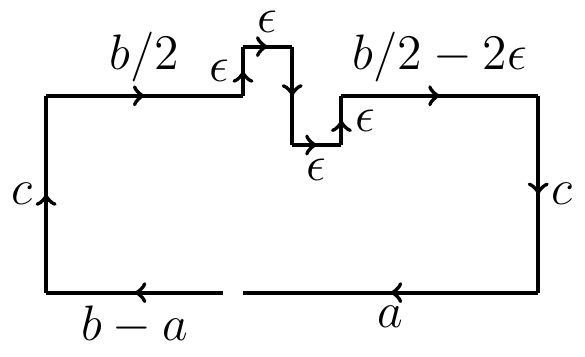}
 \caption{}
\label{Fig:P_fluctP}
\end{subfigure}
\begin{subfigure}{0.4\textwidth}
 \includegraphics[width=0.8\columnwidth]{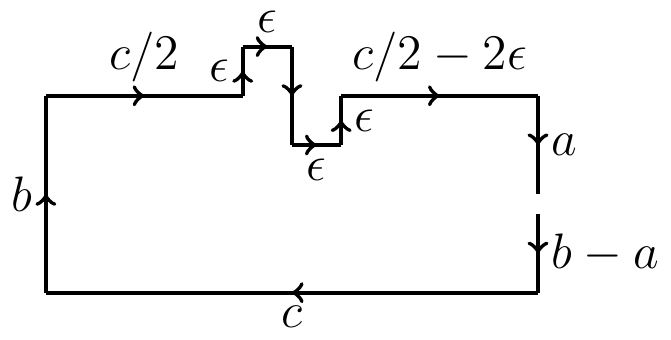}
 \caption{}
\label{Fig:X_fluctP}
\end{subfigure}

\caption{Figure showing different kinds of are-preserving fluctuations. (a) Loop starting from an arbitrary point along $X$ with fluctuations along the opposite $X$ side (b) Loop starting from an arbitrary point along $P$ with fluctuations along the adjacent $X$ side (c) Loop starting from an arbitrary point along $P$ with fluctuations along the opposite $P$ side (d) Loop starting from an arbitrary point along $X$ with fluctuations along the $P$ side}
\label{Fig:Fluctuations}
\end{figure*}

Different kinds of area-preserving deformations that are analyzed are depicted in Figure~\ref{Fig:Fluctuations}. 
The deformations have a magnitude of $\epsilon$ for loops whose dimensions are of order 1.
We give conditions for the deformations to be negligible compared to the quantum gravity signal.

Regarding the $\gamma_{0}$ case, we consider different instances of some or all of the four loops undergoing area-preserving deformations in one of the edges.
Among the different deformations, we choose the case with the largest contribution to the phase.
Under this deformation, the experimental requirements for this contribution to be less than the quantum gravity signal is 
\begin{align}
12 N_{p}^{2} \lambda_{0}^{3} k \epsilon^{3} < &\, 
\gamma_{0} \frac{\sqrt{\hbar m \omega_{m}}}{M_{p} c} \lambda_{0} N_{p}^{2} \\
\epsilon^{3} < &\, \gamma_{0} \frac{ m \omega_{m} L}{ 12 M_{p} c} 
\end{align}
Numerically, this means that $\epsilon < 10^{-4}$ for $\gamma_{0} \sim 1$.
We summarize the leading order terms in the phase in Table~\ref{Tab:Fluctuations}.

\begin{table*}
\centering
\def\arraystretch{1.5}
\begin{tabular}{| p{0.42\textwidth} | p{0.16\textwidth} | p{0.16\textwidth} |p{0.16\textwidth}|}
\hline
Final path composed of & Leading order term in $\epsilon$ & Leading order term in $\epsilon$ & Leading order term in $\epsilon$\\
& ($\gamma$ case) & ($\beta$ case) & ($\mu$ case)\\
\hline
Only one out of four loops deformed as depicted in Figure~\ref{Fig:X_fluctX} & $4 k n^3 \lambda_0^3 \epsilon^3$ &  $4 k n^3 \lambda_0^3 \epsilon^3$ & $4 k n^3 \lambda_0^3 \epsilon^3$\\
Only one out of four loops deformed as depicted in \ref{Fig:P_fluctP} respectively & $ k^{2} n^4 \lambda_0^4 \epsilon^3$ & $3 k^{2} n^4 \lambda_0^4 \epsilon^3$ & $3 k^{2} n^4 \lambda_0^4 \epsilon^3$ \\
Each of the four loops is deformed with identical $\epsilon$ along the edge opposite to the starting edge. Deformations depicted in Figures~\ref{Fig:X_fluctX} and \ref{Fig:P_fluctP} & $\frac{16}{3} k^{2} n^4 \lambda_0^4 \epsilon^3$ &  &\\
Each of the four loops is deformed but deformations arise only on the edges parallel to $X$-axis, i.e., Figures~\ref{Fig:X_fluctX} and \ref{Fig:P_fluctX} & $ \frac{842}{9} k^5 n^7 \lambda_0^7 \epsilon^3 $ & & \\
Each of the four loops is deformed but deformations arise only on the edges parallel to $P$-axis, i.e., Figures~\ref{Fig:X_fluctP} and \ref{Fig:P_fluctP} & $\frac{4}{3} k^{2} n^4 \lambda_0^4 \epsilon^3$ & & \\
\hline
In comparison, the magnitude of the signal term: & $\gamma_{0} \frac{\sqrt{\hbar m \omega_{m}}}{3 M_{p} c} \lambda_{0} n^{3} $ & $ \beta_0 \frac{\hbar \omega_m m}{3 M_p c} \lambda_0^4 n^4 $ & $\mu_{0} \frac{m^{2}}{M_{p}^{2}} \lambda_0^2 n^{2}$ \\
\hline
\end{tabular}
\caption{Summary of the leading order terms (in $\epsilon$) in the phase for different kinds of deformations.}
\label{Tab:Fluctuations}
\end{table*}

In the $\beta_{0}$ case, the corresponding requirement is 
\begin{align}
12 N_{p}^{2} \lambda_{0}^{3} k \epsilon^{3} < &\, 
\beta_0 \frac{4 \hbar \omega_m m}{3 M_p c} \lambda_0^4 N_{p}^{3} \\
\epsilon^{3} < &\, \beta_0 \frac{ \sqrt{\hbar m^{3} w_{m}^{3}} L}{9 M_p c} \lambda_{0} N_{p} 
\end{align}
Numerically, this means that $\epsilon < 10^{-6}$ for reasonable experimental parameters.
Finally, the $\mu_{0}$ case requires that
\begin{align}
12 N_{p}^{2} \lambda_{0}^{3} k \epsilon^{3} < &\, 
2 \mu_{0} \frac{m^{2}}{M_{p}^{2}} \lambda_0^2 N_{p}\\
\epsilon^{3} < &\, \mu_{0} \frac{L \sqrt{m^{3} \omega_{m}}}{6 M_{p}^{2} \lambda_{0} N_{p} \sqrt{\hbar}}, 
\end{align}
which leads to the condition that $\epsilon < 10^{2}$ or $10^{3}$ depending on the choice of reasonable experimental parameters.
This completes our analysis of the fluctuations in the phase-space paths.


\section{What if the thermal state is not a perfect thermal state?}
\label{App:Thermal}
Here we analyze the effect of imperfect preparation of the initial thermal state of the mechanical resonator and present conditions for imperfect state preparation to nullify the quantum gravity signal.
Specifically, we consider a state that is a mixture of a thermal state and a pure state
\begin{equation}
\rho = \frac{1}{1+ \epsilon} \rho_{th} + \frac{\epsilon}{1+ \epsilon} \ket{\psi}\bra{\psi}
\end{equation}
where $\ket{\psi} = \frac{1}{\sqrt{2}} \left(\ket{0} + \ket{1} \right)$, which models unwanted off-diagonal terms in the density matrix.

As usual, we evaluate the mean optical field
\begin{equation}
\braket{a} = \mathrm{Tr} \left( U^{\dagger} a U \ket{\alpha}\bra{\alpha} \otimes \rho \right).
\end{equation}
for the different quantum gravity cases. 

For the effect of these off-diagonal terms to be negligible in comparison to the QG signal, we require 
\begin{equation}
\epsilon \frac{\alpha_{0}}{\alpha'} \sin{\left( \Theta_{0} - \Phi_{QM} \right)} < \Phi_{QG},
\end{equation}
where in the $\gamma_{0}$ case we have
\begin{align}
\frac{\alpha_{0}}{\alpha'} =&\, \e^{\left( 9 |x|^{2} N_{p}^{4} + 16 |y|^{2} N_{p}^{6} + 12 \left( xy^{*} + x^{*}y \right) N_{p}^{5} \right) \left( \bar{n} - \frac{1}{2} \right)} \\
\Theta_{0} - \Phi_{QM} =&\, -\sqrt{2} k^{2} \lambda_{0}^{3} N_{p}^{2} + \frac{40}{3}\sqrt{2} k^{3} \lambda_{0}^{4} N_{p}^{3}.
\end{align} 
In the rest of this section, we derive this relation and also present the required condition for the $\beta_{0}$ and $\mu_{0}$ cases.

The unitary operator $U$ for the $\gamma_{0}$ case is given by 
\begin{align}
U = & \exp \left\{ -i w(n) + \left( x^{*}n^{3} + y^{*}n^{4} \right) a^{\dagger}_{m} 
\right. \nonumber\\
& \quad
\left. 
- \left( x n^{3} + y n^{4}\right) a_{m}\right\}
\end{align}
where 
\begin{align}
w(n) &= - \frac{40}{3} k^{3} \lambda_{0}^{5} n^{5} + 24 k^{4} \lambda_{0}^{6} n^{6} \\
x &= \left( 1- \iota \right) \frac{\sqrt{2}}{3} k^{2} \lambda_{0}^3 \\ 
y &= \left( -26 + 10 \iota \right) \frac{\sqrt{2}}{3} k^3 \lambda_{0}^4.
\end{align}

The final state of light is given by 
\begin{equation}
\braket{a} = \frac{1}{1+ \epsilon} \alpha' \e^{-i \Phi_{QM}} + \frac{\epsilon}{1+ \epsilon} \mathrm{Tr} \left( U^{\dagger} a U \ket{\alpha}\bra{\alpha} \otimes \ket{\psi} \bra{\psi} \right)
\end{equation}
where
\begin{align}
\alpha' =& \alpha
\e^{-\frac{1}{2}\left( 9 |x|^{2} N_{p}^{4} + 16 |y|^{2} N_{p}^{6} \right)}
\nonumber\\
& 
\times
\e^{-\left( 9 |x|^{2} N_{p}^{4} + 16 |y|^{2} N_{p}^{6} + 12 \left( xy^{*} + x^{*}y \right) N_{p}^{5} \right) \bar{n}} 
\end{align}
and 
\begin{equation}
\Phi_{QM} = \frac{1}{2i}\left( x^{*}y - x y^{*}\right) N_{p}^{6} 
- \frac{200}{3} k^{3} \lambda_{0}^{5} N_{p}^{4}
+ 144 k^{4} \lambda_{0}^{6} N_{p}^{5}
\end{equation}
as can be seen from calculations in Appendix~\ref{App:MeanField4Displacement}.

We now evaluate the second part of the expression $\braket{a_{0}} = \mathrm{Tr} \left( U^{\dagger} a U \ket{\alpha}\bra{\alpha} \otimes \ket{\psi} \bra{\psi} \right)$. 
Performing calculations similar to that in Appendix~\ref{App:MeanField4Displacement} we see that 
\begin{align}
U^{\dagger} a U
= & \,
\e^{-\frac{1}{2}\left( x^{*}y - x y^{*}\right)\left( n^{6} + 15 n^{5} + 33n^{4} +35 n^{3} +21 n^{2} + 7n + 1 \right)}
\nonumber\\
& 
\times
\e^{i \frac{40}{3} k^{3} \lambda_{0}^{5} \left( 5n^{4} + 10n^{3} + 10n^{2} + 5n+ 1 \right)} 
\nonumber\\
& 
\times
\e^{-i 24 k^{4} \lambda_{0}^{6} \left( 6 n^{5} + 15n^{4} +20 n^{3} +15 n^{2} + 6n + 1 \right) } 
\nonumber\\
& 
\times
\e^{\left( x^{*}a^{\dagger}_{m} - x a_{m} \right) \left( 3n^{2} + 3n +1 \right) }
\nonumber\\
& 
\times
\e^{\left( y^{*}a^{\dagger}_{m} - y a_{m} \right) \left( 4n^{3} + 6n^{2} + 4n+ 1 \right) } 
a.
\end{align}
Therefore,
\begin{align}
\braket{a_{0}} = & \, 
\sum_{n=0}^{\infty} 
\alpha
\braket{\alpha | n} \braket{n|\alpha}
\e^{i \frac{40}{3} k^{3} \lambda_{0}^{5} \left( 5n^{4} + 10n^{3} + 10n^{2} + 5n+ 1 \right)} 
\nonumber\\
& 
\times
\e^{-i 24 k^{4} \lambda_{0}^{6} \left( 6 n^{5} + 15n^{4} +20 n^{3} +15 n^{2} + 6n + 1 \right) } 
\nonumber\\
& 
\times
\e^{-\frac{1}{2}\left( x^{*}y - x y^{*}\right)\left( n^{6} + 15 n^{5} + 33n^{4} +35 n^{3} +21 n^{2} + 7n + 1 \right)}
\nonumber\\
& 
\times
\bra{\psi}
\e^{\left( x^{*}a^{\dagger}_{m} - x a_{m} \right) \left( 3n^{2} + 3n +1 \right) }
\nonumber\\
& 
\times
\e^{\left( y^{*}a^{\dagger}_{m} - y a_{m} \right) \left( 4n^{3} + 6n^{2} + 4n+ 1 \right) } 
\ket{\psi}.
\end{align}
We define the variables
\begin{align}
\upsilon &= y \left( 4n^{3} + 6n^{2} + 4n+ 1\right) \\
\chi &= x \left( 3n^{2} + 3n +1 \right).
\end{align}
and denote the displaced Fock state $\e^{\left( \upsilon^{*}a^{\dagger}_{m} - \upsilon a_{m} \right)} \ket{m}$ as $\ket{\upsilon^{*}, m}$. 
We rewrite 
$$\bra{\psi} \e^{\left( x^{*}a^{\dagger}_{m} - x a_{m} \right) \left( 3n^{2} + 3n +1 \right) } \e^{\left( y^{*}a^{\dagger}_{m} - y a_{m} \right) \left( 4n^{3} + 6n^{2} + 4n+ 1 \right) } \ket{\psi}$$ as
\begin{align}
& \bra{\psi}
\e^{\left( \chi^{*}a^{\dagger}_{m} - \chi a_{m} \right)}\e^{\left( \upsilon^{*}a^{\dagger}_{m} - \upsilon a_{m} \right)} 
\ket{\psi} = 
\frac{1}{2} \left\{ \braket{-\chi^{*},0 | \upsilon^{*},0} \right.
\nonumber \\
& \quad \left.
+ \braket{-\chi^{*},0 | \upsilon^{*},1} 
+ \braket{-\chi^{*},1 | \upsilon^{*},0} 
+ \braket{-\chi^{*},1 | \upsilon^{*},1} \right\}
\end{align}

Using the formula for the overlap of two displaced Fock states from \cite{Wunsche1991}
\begin{align}
& \braket{-\chi^{*} , m | \upsilon^{*}, n}  = \braket{-\chi^{*} | \upsilon^{*}} \sqrt{m!n!} 
\nonumber \\
& \qquad \times
\sum_{j=0}^{\mathrm{min}(m,n) } \frac{ \left( \upsilon^{*} + \chi^{*}\right)^{m-j} \left( -\chi - \upsilon \right)^{n-j} }{j! \left( m-j \right)! \left( n-j \right)! }
\end{align}
where
\begin{equation}
 \braket{-\chi^{*} | \upsilon^{*}} = \exp \left\{ -\chi \upsilon^{*} -\frac{1}{2}\left( |\chi|^{2} + |\upsilon|^{2} \right) \right\}
\end{equation}
we can evaluate the expression.

\begin{align}
\bra{\psi}
& \e^{\left( \chi^{*}a^{\dagger}_{m} - \chi a_{m} \right)}\e^{\left( \upsilon^{*}a^{\dagger}_{m} - \upsilon a_{m} \right)}
\ket{\psi} = 
\frac{1}{2}\braket{-\chi^{*} | \upsilon^{*}}
\left\{ 1+ \chi^{*} 
\right.
\nonumber \\
& \, 
\left.
+ \upsilon^{*} -\chi -\upsilon 
+1 - |\chi|^{2} - |\upsilon|^{2} -\chi \upsilon^{*} - \chi^{*}\upsilon \right\}
\end{align}

Also note that the other terms that are in the expression for $\braket{a}$ are given by
\begin{equation}
\braket{\alpha|n} \braket{n|\alpha} = \e^{-|\alpha|^{2}} \frac{|\alpha|^{2n}}{n!}.
\end{equation}
Substituting all the above expressions in the expression for $\braket{a_{0}}$, we get
\begin{align}
\braket{a_{0}} = & \, 
\sum_{n=0}^{\infty} 
\alpha
\e^{-|\alpha|^{2}} \frac{|\alpha|^{2n}}{n!}
\e^{i \frac{40}{3} k^{3} \lambda_{0}^{5} \left( 5n^{4} + 10n^{3} + 10n^{2} + 5n+ 1 \right)} 
\nonumber\\
& 
\times
\e^{-i 24 k^{4} \lambda_{0}^{6} \left( 6 n^{5} + 15n^{4} +20 n^{3} +15 n^{2} + 6n + 1 \right) } 
\nonumber\\
& 
\times
\e^{-\frac{1}{2}\left( x^{*}y - x y^{*}\right)\left( n^{6} + 15 n^{5} + 33n^{4} +35 n^{3} +21 n^{2} + 7n + 1 \right)}
\nonumber\\
& 
\times
\e^{ -xy^{*} \left( 3n^{2} + 3n +1 \right) \left( 4n^{3} + 6n^{2} + 4n+ 1\right)} 
\nonumber\\
& 
\times
\e^{-\frac{1}{2}\left( |x|^{2}\left( 3n^{2} + 3n +1 \right)^{2} + |y|^{2}\left( 4n^{3} + 6n^{2} + 4n+ 1\right)^{2} \right) }
\nonumber\\
& 
\times
\frac{1}{2} \left\{ 1
+ \left( x^{*} - x \right) \left( 3n^{2} + 3n +1 \right) 
\right.
\nonumber \\
& \, 
+ \left( y^{*} - y \right) \left( 4n^{3} + 6n^{2} + 4n+ 1\right) 
+1 
\nonumber \\
& \, 
- |x|^{2}\left( 3n^{2} + 3n +1 \right)^{2} - |y|^{2}\left( 4n^{3} + 6n^{2} + 4n+ 1\right)^{2} 
\nonumber \\
& \, 
- \left( x^{*}y + x y^{*}\right)\left( 12 n^{5} + 30 n^{4} +34 n^{3} 
\right. \nonumber \\
& \, 
\left. \left.
+21 n^{2} + 12 n + 1 \right)
 \right\}.
\end{align}
The expression can be approximated (to leading order in $N_{p}$) to be
\begin{align}
\braket{a_{0}} & = 
\alpha
\e^{-\frac{1}{2}\left( 9 |x|^{2} N_{p}^{4} + 16 |y|^{2} N_{p}^{6} \right)}
\nonumber\\
& 
\times
\e^{-\frac{1}{2}\left( x^{*}y - x y^{*}\right) N_{p}^{6} 
+ i \frac{200}{3} k^{3} \lambda_{0}^{5} N_{p}^{4}
-i 144 k^{4} \lambda_{0}^{6} N_{p}^{5}
}
\nonumber\\
& 
\times
\frac{1}{2} \left\{ 1+ 3 \left( x^{*} - x \right) N_{p}^{2}  + 4 \left( y^{*} - y \right) N_{p}^{3} - 9 |x|^{2} N_{p}^{4}  \right.
\nonumber \\
& \, 
\left.  +1 - 16 |y|^{2} N_{p}^{6} - 12 \left( x^{*}y + x y^{*}\right) N_{p}^{5}
 \right\}.
\end{align}

If $\braket{a_{0}}$ is given by 
\begin{equation}
\braket{a_{0}} = \alpha_{0} \e^{-i \Theta_{0}}, 
\end{equation}
the new amplitude is 
\begin{equation}
\alpha_{0} \approx \alpha
\e^{-\left( 9 |x|^{2} N_{p}^{4} + 16 |y|^{2} N_{p}^{6} + 6 \left( x^{*}y + x y^{*}\right) N_{p}^{5}\right)}
\end{equation}
and the new phase is
\begin{align}
\Theta_{0} \approx & \frac{1}{2i}\left( x^{*}y - x y^{*}\right) N_{p}^{6} 
- \frac{200}{3} k^{3} \lambda_{0}^{5} N_{p}^{4}
+ 144 k^{4} \lambda_{0}^{6} N_{p}^{5}
\nonumber \\
& \, 
- \frac{1}{2i} \left\{ 3 \left( x^{*} - x \right) N_{p}^{2}  + 4 \left( y^{*} - y \right) N_{p}^{3} \right\}
\end{align}
which on substituting with $x$ and $y$ gives
\begin{align}
\Theta_{0} \approx &
- \frac{32}{9} k^{5} \lambda_{0}^{7} N_{p}^{6} 
- \frac{200}{3} k^{3} \lambda_{0}^{5} N_{p}^{4}
+ 144 k^{4} \lambda_{0}^{6} N_{p}^{5}
\nonumber \\
& \, 
- \sqrt{2} k^{2} \lambda_{0}^{3} N_{p}^{2} 
+ \frac{40 \sqrt{2}}{3} k^{3} \lambda_{0}^{4} N_{p}^{3}.
\end{align}

Putting the two equations together, the final state of light is given by
\begin{equation}
\braket{a} = \frac{1}{1+ \epsilon} \alpha' \e^{-i \Phi_{QM}} + \frac{\epsilon}{1+ \epsilon} \alpha_{0} \e^{-i \Theta_{0}}. 
\end{equation}
In the remainder of this section, we calculate the effective amplitude and phase of the light. 
The mean field is simplified to 
\begin{equation}
\braket{a} = \frac{1}{1+ \epsilon} \alpha' \e^{-i \Phi_{QM}}\left( 1 + \epsilon \frac{\alpha_{0}}{\alpha'} \e^{-i \left( \Theta_{0} - \Phi_{QM} \right)} \right).
\end{equation}
Define
\begin{equation}
 \epsilon \frac{\alpha_{0}}{\alpha'} \e^{-i \left( \Theta_{0} - \Phi_{QM} \right)} = 
 r \e^{{i \phi}}.
\end{equation}
Expressing $1+ r \e^{{i \phi}}$ in the polar form, we have
\begin{equation}
1+ r \e^{{i \phi}} = \sqrt{1+r^{2} + 2 r \cos{\phi}} \, \e^{i \tan^{-1} \left( \frac{r \sin{\phi}}{1+r \cos{\phi}} \right)}
\end{equation}
which to first order in r is (first order in $\epsilon$)
\begin{equation}
1+ r \e^{{i \phi}} = \left(1+ r \cos{\phi} \right) \, \e^{i r \sin{\phi}}. 
\end{equation}
Therefore, the mean field is given by
\begin{align}
\braket{a} =& \frac{1}{1+ \epsilon} \alpha' 
\left(1+ \epsilon \frac{\alpha_{0}}{\alpha'} \cos{\left( \Theta_{0} - \Phi_{QM} \right)} \right) \e^{-i \Phi_{QM}}
\nonumber \\
& \, 
\e^{-i \epsilon \frac{\alpha_{0}}{\alpha'} \sin{\left( \Theta_{0} - \Phi_{QM} \right)}},
\end{align}
where the respective amplitude and phase of the output light are given by 
\begin{align}
\alpha' =&\, \alpha
\e^{-\frac{1}{2}\left( 9 |x|^{2} N_{p}^{4} + 16 |y|^{2} N_{p}^{6} \right)}
\nonumber \\
& \times 
\e^{-\left( 9 |x|^{2} N_{p}^{4} + 16 |y|^{2} N_{p}^{6} + 12 \left( xy^{*} + x^{*}y \right) N_{p}^{5} \right) \bar{n}},\\
\frac{\alpha_{0}}{\alpha'} =&\, \e^{\left( 9 |x|^{2} N_{p}^{4} + 16 |y|^{2} N_{p}^{6} + 12 \left( xy^{*} + x^{*}y \right) N_{p}^{5} \right) \left( \bar{n} - \frac{1}{2} \right)} 
\end{align} 
and
\begin{align}
\Phi_{QM} =&\, 
- \frac{200}{3} k^{3} \lambda_{0}^{5} N_{p}^{4}
+ 144 k^{4} \lambda_{0}^{6} N_{p}^{5}
- \frac{32}{9} k^{5} \lambda_{0}^{7} N_{p}^{6}
,\\
\Theta_{0} - \Phi_{QM} =&\, -\sqrt{2} k^{2} \lambda_{0}^{3} N_{p}^{2} + \frac{40}{3}\sqrt{2} k^{3} \lambda_{0}^{4} N_{p}^{3}.
\end{align}

Note that the correction to the phase is given by $ \epsilon \frac{\alpha_{0}}{\alpha'} \sin{\left( \Theta_{0} - \Phi_{QM} \right)}$.
Thus, for imperfect preparation to have no significant impact on the results, we require 
\begin{equation}
\epsilon \frac{\alpha_{0}}{\alpha'} \sin{\left( \Theta_{0} - \Phi_{QM} \right)} < \Phi_{QG}.
\end{equation}

Analogous calculations for the $\mu_{0}$ case yield
\begin{align}
\alpha' = & \alpha
\e^{-\frac{1}{2}\left( 4 |x|^{2} N_{p}^{2} + 9 |y|^{2} N_{p}^{4} \right)}
\nonumber \\
& \, 
\e^{-\left( 4 |x|^{2} N_{p}^{2} + 9 |y|^{2} N_{p}^{4} + 6 \left( xy^{*} + x^{*}y \right) N_{p}^{3} \right) \bar{n}} ,
\end{align} 
\begin{equation}
\Phi_{QM} = 2 \lambda_{0}^{2} N_{p} - 2k^{3} \lambda_{0}^{5} N_{p}^{4}, 
\end{equation}
\begin{equation}
\frac{\alpha_{0}}{\alpha'} = 
\e^{\left( 4 |x|^{2} N_{p}^{2} + 9 |y|^{2} N_{p}^{4} + 6 \left( xy^{*} + x^{*}y \right) N_{p}^{3} \right) \left( \bar{n} - \frac{1}{2} \right)},
\end{equation}
and
\begin{equation}
\Theta_{0} - \Phi_{QM} = - 2 \sqrt{2} k \lambda_{0}^{2} N_{p} + \frac{3}{\sqrt{2}} k^{2} \lambda_{0}^{3} N_{p}^{2}
\end{equation}
for 
\begin{align}
x &= \left( -1- i \right) \sqrt{2} k \lambda_{0}^2 \\ 
y &= \left( -1 + i \right) \frac{1}{\sqrt{2}} k^2 \lambda_{0}^3. 
\end{align}

Similarly for the $\beta_{0}$ case, we have 
\begin{align}
\alpha' =& \, \alpha
\exp \left\{-\frac{1}{2}\left( 4 |x|^{2} N_{p}^{2} + 9 |y|^{2} N_{p}^{4} + 16 |z|^{2} N_{p}^{6} \right) \right\}
\nonumber\\
& 
\times
\exp \left\{-\left( 4 |x|^{2} N_{p}^{2} + 9 |y|^{2} N_{p}^{4} + 16 |z|^{2} N_{p}^{6} 
\right. \right. \nonumber\\
& \left. \left.
+ 6 \left( xy^{*} + x^{*}y \right) N_{p}^{3} + 12 \left( yz^{*} + y^{*}z \right) N_{p}^{5} 
\right. \right. \nonumber\\
& \left. \left.
+ 8 \left( xz^{*} + x^{*}z \right) N_{p}^{4} \right) \bar{n}\right\}, 
\end{align}
\begin{equation}
\Phi_{QM} = 2 \lambda_{0}^{2} N_{p} 
+ 6 k^{2} \lambda_{0}^{2} N_{p} 
- 2k^{3} \lambda_{0}^{5} N_{p}^{4}
- 4 k^{5} \lambda_{0}^{7} N_{p}^{6}.
\end{equation}
\begin{align}
\frac{\alpha_{0}}{\alpha'} = &
\exp \left\{ \left( 4 |x|^{2} N_{p}^{2} + 9 |y|^{2} N_{p}^{4} + 16 |z|^{2} N_{p}^{6} 
\right. \right. \nonumber\\
& \left. \left.
+ 6 \left( xy^{*} + x^{*}y \right) N_{p}^{3} + 12 \left( yz^{*} + y^{*}z \right) N_{p}^{5} 
\right. \right. \nonumber\\
& \left. \left.
+ 8 \left( xz^{*} + x^{*}z \right) N_{p}^{4} \right) \left( \bar{n} - \frac{1}{2} \right) \right\}, 
\end{align}
and
\begin{equation}
\Theta_{0} - \Phi_{QM} = - 2 \sqrt{2} k \lambda_{0}^{2} N_{p} + \frac{3}{\sqrt{2}} k^{2} \lambda_{0}^{3} N_{p}^{2} + 8 \sqrt{2} k^{3} \lambda_{0}^4 N_{p}^{3}
\end{equation}
for 
\begin{align}
x &= \left( -1- i \right) \sqrt{2} k \lambda_{0}^2 \\ 
y &= \left( -1 + i \right) \frac{1}{\sqrt{2}} k^2 \lambda_{0}^3\\
z &= \left( 1+ i \right) 2\sqrt{2} k^{3} \lambda_{0}^4 . 
\end{align}

Thus, the state preparation should be such that the phase contribution $\epsilon \frac{\alpha_{0}}{\alpha'} \sin{\left( \Theta_{0} - \Phi_{QM} \right)}$ of the off-diagonal terms is less than the quantum gravity signal.

\section{Open problem: Accuracy of the assumptions made in the calculations}
\label{App:Convergence}
In Section~\ref{Sec:Results} we calculated the phase acquired by light after the action of the suggested operator~\eqref{Eq:GammaFourLoop}.
Similar calculations are detailed in~Appendix~\ref{App:MeanField4Displacement}.
In the calculation of the phase, we make several approximations. 
We note that this is not a deficiency of our approach but also arises implicitly in Ref.~\cite{Pikovski2012} where the effect of the truncation was however not estimated.
In this section, we describe the approximations made and discuss their validity.

First, we recall the assumptions made in simplifying the unitary operator.
In order to calculate the phase, we first need to express the unitary operator, which is a product~\eqref{Eq:ProductU} of exponentials of operators, as a single exponential of operators~\eqref{Eq:Uexact}. 
This simplification cannot be performed exactly for an arbitrary Hamiltonian.
So, we need to truncate the Hamiltonian~\eqref{Eq:App_HXHP} to a finite order in $k$.
This is our first approximation.
The second approximation is choosing a finite order in BCH formula based on available computation resources.
The unitary operator thus calculated has many terms in the exponent. 
We calculate the phase contribution from only the terms larger than the minimum uncertainty and ignore the rest to obtain the approximate unitary operator~\eqref{Eq:UTrunc}, thus making our third approximation. 
The final approximation made is the saddle-point approximation, which is employed in going from Equation~\eqref{Eq:aExactSum} to Equation~\eqref{Eq:aApprox}.

The exact (or general) form of the phase from the unitary operator calculated to an arbitrary order of BCH formula or $k$ is not known.
So, it is difficult to prove convergence of the phase rigorously.
We, therefore, try to check the validity of the assumptions heuristically. 
One approach to check the validity of truncation in the BCH order is to fix a specific order in $k$ (like $k=2$) and calculate the phase contribution from each order in BCH. 
A sufficient condition for the validity of our assumptions is that these phase contributions fall off quickly with increasing BCH order.
However, we were not able to calculate the phase for each BCH order exactly. 
This is because even for a finite number of terms in each BCH order, there can be infinitely many phase terms as illustrated in \ref{Sec:InfinitePhase}.
Checking the validity of truncation in the $k$ order has the same challenge of not being able to calculate the phase.
Also, since the phase cannot be calculated exactly even for a few terms in the unitary operator for a given BCH and $k$ order, we cannot comment on the validity of truncating the unitary operator.
We leave a proof of the validity of these assumptions as an open problem.

We instead give evidence to the validity of the approximations of calculating the unitary operator to a given BCH and $k$ order.
We give evidence that the term with the largest phase contribution from each BCH order drops off geometrically, even though we can make no statement about the sum of all terms from that BCH order.
Consider simplifying the expression $\e^{\iota H_{X}} \e^{\iota H_{P}}$ for
\begin{align}
\begin{split}
H_{X} =& \, n \lambda_0 \left( X - k X^2 + k^{2} X^3 - \dots \right) \\
H_{P} =& \, n \lambda_0 \left( P - k P^2 + k^{2} P^3 - \dots \right).
\label{Eq:HxHp}
\end{split}
\end{align}
In the final simplification, we see that the phase contribution from the terms constant in $X$ and $P$ is larger than the phase contribution from the non-constant terms.
This is because the terms dependent on the mechanical modes (terms with $X$ and $P$) only contribute through their commutators while the constant terms contribute directly as can be seen in the calculations in Appendix~\ref{App:MeanField4Displacement}.
So, for small coefficients ($k n \lambda_{0} < 1$), the largest contribution is from the constant term.

The constant term from first order in BCH, $\left[ H_{X}, H_{P} \right]$, comes from $\left[ X, P \right]$ and is therefore of the order $n^{2} \lambda_{0}^{2}$. 
The constant term from second order comes from terms like $\left[ X, \left[ X, P^{2} \right] \right]$ and $\left[ P, \left[ P, X^{2} \right] \right]$. 
We see that these terms have coefficients of the order of $k n^{3} \lambda_{0}^{3}$.
Similarly, the constant term from BCH order $m$ is of the order of $k^{m-2} n^{m} \lambda_{0}^{m}$. 
So, the largest term in each BCH order falls of geometrically.
The phase contribution from such a term is of the order of $k^{m-2} \lambda_{0}^{m} N_{p}^{m-1}$. 
If we assume that the sum of the rest of the terms is negligible, we need only to calculate up to BCH order $m$ such that the phase is less than the minimum uncertainty $\Delta \Phi_{T}$. 
That is, 
\begin{equation}
k^{m-2} \lambda_{0}^{m} N_{p}^{m-1} < \frac{1}{2\sqrt{N_{p} N_{r}}}.
\end{equation}
For the $\gamma_{0}$ experimental parameters, we estimate that this condition is satisfied for $m=6$. 
So we need to calculate up to BCH order 6 and $k$ order 4.
Similarly, $m=28$ in the $\beta_{0}$ case and $m=5$ in the $\mu_{0}$ case suffice.
\subsection{Infinite number of phase terms from unitary operator}
\label{Sec:InfinitePhase}

Consider the case when the unitary operator is given by 
\begin{equation}
U = \e^{\chi n^{2} + \upsilon n^{3}}
\label{Eq:NonlinU}
\end{equation}
where $\chi$ is linear and $\upsilon$ quadratic in $a_{m}$ and $a^{\dag}_{m}$.
For example
\begin{equation}
\begin{split}
\chi = x^{*} a^{\dag}_{m} - x a_{m} \\
\upsilon = y^{*} a^{\dag 2}_{m} - y a^{2}_{m}.
\label{Eq:NonlinXY}
\end{split}
\end{equation}

The final quantity to be calculated is the mean field of light which is
\begin{equation}
\braket{a} = \mathrm{Tr} \left( U^{\dagger} a U \ket{\alpha}\bra{\alpha} \otimes \rho^{th}_{m} \right).
\end{equation}
The first step in this calculation is to simplify $U^{\dagger} a U$ and express it as $O a$ where $O$ is an operator only dependent on $n$.
$a$ acts on $\ket{\alpha}$ to give $\alpha \ket{\alpha}$. 
The operator $a$ can then be removed from the trace.
The mean field is then given by 
\begin{equation}
\braket{a} = \alpha \mathrm{Tr} \left( O \ket{\alpha}\bra{\alpha} \otimes \rho^{th}_{m} \right).
\end{equation}

We show that the approach that works in simplifying $U^{\dagger} a U$ used in the calculations of Appendix~\ref{App:MeanField4Displacement} does not work in the case where the unitary operator is of the form given by Equations~\eqref{Eq:NonlinU} and~\eqref{Eq:NonlinXY}.

This approach relies on splitting $U$ into a product of exponentials each containing one power of $n$ as is done while going from Equation~\eqref{Eq:Split1} to~\eqref{Eq:Split2}.
Using the Zassenhaus formula
\begin{align}
& \e^{(\chi+\upsilon)} = \e^{\chi}
~\e^{\upsilon}
~\e^{-{\frac {1}{2}}[\chi,\upsilon]}
~\e^{{\frac {1}{6}}(2[\upsilon,[\chi,\upsilon]]+[\chi,[\chi,\upsilon]])}
\nonumber \\ & \quad \times
~\e^{{-\frac {1}{24}}([[[\chi,\upsilon],\chi],\chi]+3[[[\chi,\upsilon],\chi],\upsilon]+3[[[\chi,\upsilon],\upsilon],\upsilon])}
\dots
\end{align}
we see that the expansion does not truncate due to $\upsilon$ being quadratic in $a_{m}$ and $a^{\dag}_{m}$.
Terms of the form $[\upsilon,\chi]$, $[\upsilon, [\upsilon,\chi]]$, $[\upsilon, [\upsilon,[\upsilon,\chi]]]$ and so on are non-zero and linear functions of $a_{m}$ and $a^{\dag}_{m}$.
Terms of the form $[\chi, [\upsilon,\chi]]$, $[\chi, [\upsilon, [\upsilon,\chi]]]$, $[\chi, [\upsilon, [\upsilon,[\upsilon,\chi]]]]$ are non-zero and functions of $n$ alone, not $a_{m}$ or $a^{\dag}_{m}$.
So, even with a small number of terms in the unitary operator, we cannot calculate the phase of the state of light exactly.

\section{Calculation details of the $\beta_{0}$ and $\mu_{0}$ case}
\label{App:BetaAndMu}

\subsection{Challenges in Pikovski \emph{et al.}~analysis}

Here we present an analysis of the cases of quantum gravity parameters $\beta_{0}$ and $\mu_{0}$ similar to that of Section~\ref{Sec:Problems}, which deals with the $\gamma_{0}$ case.

\subsubsection{$\beta_{0}$ case}
Here, the quantum gravity signal is given by 
\begin{equation}
\Phi_{QG} =\beta_0 \kappa' \lambda_0^4 N_p^3 \quad \text{for} \quad \kappa' := \frac{4 \hbar m \omega}{3 M_p c}.
\end{equation}
The expression for $\beta_{0}$ now reads
\begin{equation}
\beta_0 = \frac{1}{ \kappa' \lambda_0^4} \left( \frac{\Phi_T}{N_p^3}\right) -  \frac{2}{\kappa' \lambda_{0}^{2} N_{p}^{2}} + \frac{6 k}{\kappa' \lambda_{0} N_{p}} - \frac{16 k^{2}}{\kappa'}
\end{equation}
and its standard deviation is given by
\begin{align}
& \left( \Delta \beta_0 \right)^2 = 
\left( \frac{1}{ \kappa' \lambda_0^4 N_p^3} \right)^2 \left(\Delta \Phi_{T} \right)^2 
+ \left( - \frac{3 \Phi_T}{ \kappa' \lambda_0^4 N_p^4} 
\right. \nonumber \\
& \quad \left.
+  \frac{4}{\kappa' \lambda_{0}^{2} N_{p}^{3}} 
- \frac{6 k}{\kappa' \lambda_{0} N_{p}^{2}} \right)^2 \left( \Delta N_p \right)^2
\end{align}
for one run of the experiment.
To estimate the precision, we substitute for $\Phi_{T}$ and evaluate $\left( \Delta \beta_0 \right)^2$ for $\beta_{0} \sim 0$. We obtain that for a precision of $\left( \Delta \beta_0 \right)^2 \sim 1$, we need to perform the experiment $N_{r} = 10^{19}$ ($10^{25}$) times in the quantum-noise-limited (classical-noise-limited) scheme, which is much less feasible than the $10^{6}$ required experimental runs claimed in~\cite{Pikovski2012}.

\subsubsection{$\mu_{0}$ case}
The quantum gravity signal is rewritten as
\begin{equation}
\Phi_{QG} = \mu_0 \kappa'' \lambda_0^2 N_p \quad \text{for} \quad \kappa'' := 2\frac{m^{2}}{M_{p}^{2}}.
\end{equation}
The expression for $\mu_{0}$ is given by
\begin{equation}
\mu_0 = \frac{1}{ \kappa'' \lambda_0^2} \left( \frac{\Phi_T}{N_p}\right) -  \frac{2}{\kappa''} + \frac{6 k \lambda_{0} N_{p}}{\kappa''} - \frac{16 k^{2} \lambda_{0}^{2} N_{p}^{2}}{\kappa''}.
\end{equation}
and its variance is
\begin{align}
& \left( \Delta \mu_{0} \right)^2 =  \left( \frac{1}{ \kappa'' \lambda_0^2 N_p} \right)^2 \left(\Delta \Phi_{T} \right)^2 
+ \left( - \frac{ \Phi_T}{ \kappa'' \lambda_0^2 N_p^2}
+ \frac{6 k \lambda_{0}}{\kappa''} 
\right. \nonumber \\
& \quad \left.
- \frac{32 k^{2} \lambda_{0}^{2} N_{p}}{\kappa''} \right)^2 \left( \Delta N_p \right)^2
\end{align}
for one run of the experiment.
Substituting for $\Phi_{T}$, and assuming $\mu_{0} \sim 0$, the variance is $\left( \Delta \mu_0 \right)^2 = 10^{5}$ in both schemes. So, to have $\left( \Delta \mu_0 \right)^2 \sim 1$, we need to perform the experiment $N_{r} = 10^{5} $ times as opposed to $O(1)$ times~\cite{Pikovski2012}.

\subsection{Results for $\beta_{0}$ and $\mu_{0}$ cases}
Here we present results for $\beta_{0}$ and $\mu_{0}$ cases similar to those of Section~\ref{Sec:Results} for the $\gamma_{0}$ case.
Specifically, we calculate the expected phase for the $\beta_{0}$ and $\mu_{0}$ using unitary operators similar to that detailed in~Appendix~\ref{App:MeanField4Displacement}.

\subsubsection{$\beta_{0}$ case}
\label{Sec:BetaResult}

Now we present the analysis for the $\beta_{0}$ case.
In this case, we suggest two possible solutions $U_{\beta_{0},1}$ and $U_{\beta_{0},2}$ with different advantages and disadvantages. 
The first solution is described below.

\textbf{First solution} --
This path in phase space is identical to the one path suggested for the $\gamma_{0}$ case. 
We perform similar calculations as in the $\gamma_{0}$ case to evaluate $U_{\beta_{0},1}$.
Each of the individual loops was evaluated to sixth order in the BCH formula and the composition of the four loops was evaluated to third order in BCH formula to obtain the phase
\begin{align}
\Phi_{T} = & \, - \frac{40}{9} \beta \lambda_0^4 N_{p}^{3} 
- \frac{200}{3} k^3 \lambda_{0}^5 N_{p}^4
+ 144 k^4 \lambda_{0}^6 N_{p}^5
\nonumber \\
& \,
+ \frac{1624}{3} k^5 \lambda_{0}^7 N_{p}^6 
- \frac{99680}{27} k^6 \lambda_{0}^8 N_{p}^7
\nonumber \\
& \,
- 3116 k^{7} \lambda_{0}^9 N_{p}^8
+ \dots.
\label{Eq:BetaPhiT}
\end{align}

The advantage of this solution is that the total number of runs required decreases by a few orders of magnitude.
However, a major disadvantage of this four-loop path is that the assumptions made in the above evaluation of $U_{\beta_{0},1}$ (and the acquired phase thereof) are not controlled. 
In more detail, increasing the BCH order from 5 to 6 while evaluating the composition of the four loops leads to an additional contribution to the phase that is larger than the quantum gravity signal.
Thus, there is no evidence that the phase obtained from the BCH approximations for higher than 6 orders is insignificant.
In summary, the four-loop path is infeasible for estimating $\beta_{0}$ requires overcoming potential issues with the convergence of the expected phase.
Instead, we suggest a different solution.

\begin{figure}[h]\centering
 \includegraphics[height=0.3\columnwidth]{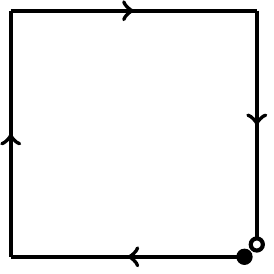}
\caption{$U_{\beta_{0},2}$: The shape of the path that reduces some of the QM contribution for the $\beta_{0}$ commutator.}
\label{Fig:UBeta2}
\end{figure}

\textbf{Second solution} --
This path in phase space is composed of only one rectangular loop like the original~\cite{Pikovski2012} but starting at a different vertex of the rectangle.
The path is given by $U_{\beta_{0},2}$.
\begin{equation}
U_{\beta_{0},2} = 
\e^{- \iota H_{X}} 
\e^{-\iota H_{P}} 
\e^{\iota H_{X}}
\e^{\iota H_{P}}
\end{equation}
The path is depicted in Figure~\ref{Fig:UBeta2}.

Performing calculations similar to those in Equations~\eqref{Eq:Optical field}--\eqref{Eq:PhiQG}, we calculate the phase of the measured light $\Phi_{T}$ to be 
\begin{equation}
\Phi_{T} = \frac{4}{3}\beta \lambda_{0}^4 N_{p}^{3}
+ 2 \lambda_{0}^{2} N_{p} 
- 2k^{3} \lambda_{0}^{5} N_{p}^{4}
+ 35 k^{4} \lambda_{0}^{6} N_{p}^{5}
- 4 k^{5} \lambda_{0}^{7} N_{p}^{6}.
\label{Eq:AppPhiTBeta0}
\end{equation}

Since the shape of the loop remains the same as in the original case, the largest contribution to the quantum mechanical phase remains the same. 
However, the second-largest term is reduced by 2 orders of magnitude by starting from the different vertex in the loop, which simplifies the phase calculations substantially. 

Now we estimate the number of experimental runs required for the precise estimation of $\beta_{0}$.
From Equation~\eqref{Eq:AppPhiTBeta0}, $\beta_{0}$ is determined from the total measured phase $\Phi_{T}$ using the relation
\begin{align}
\beta_0 = & \frac{3}{4 \kappa' \lambda_0^4} \left( \frac{\Phi_T}{N_p^3}\right) 
- \frac{3}{2 \kappa' \lambda_{0}^{2} N_{p}^{2}}
+ \frac{3 k^{3} \lambda_{0} }{2 \kappa'} N_{p}
- \frac{105 k^{4} \lambda_{0}^{2} }{4 \kappa'} N_{p}^{2}
\nonumber \\
& \,
+ \frac{3 k^{5} \lambda_{0}^{3} }{\kappa'} N_{p}^{3}
\end{align}
for 
\begin{equation}
\kappa' := \frac{ \hbar m \omega}{M_p c}.
\end{equation}
The uncertainty in $\beta_0$ for one run of the experiment is given by
\begin{align}
& \left( \Delta \beta_0 \right)^2  = 
 \left( - \frac{9 \Phi_T}{ 4 \kappa \lambda_0^4 N_p^4} 
+ \frac{3}{\kappa' \lambda_{0}^{2} N_{p}^{3}} 
+ \frac{3 k^{3} \lambda_{0} }{2 \kappa'} 
\right. \nonumber \\
& \quad \left.
- \frac{105 k^{4} \lambda_{0}^{2} N_{p}}{2 \kappa'}
\right)^{2} 
\left( \Delta N_p \right)^2
+ \left( \frac{3}{ 4 \kappa \lambda_0^4 N_p^3} \right)^2 \left(\Delta \Phi_{T} \right)^2 
\label{Eq:BetaUncertainty}
\end{align}

With the following experimental parameters, as suggested in~\cite{Pikovski2012}
\begin{align*}
N_p &= 10^{14}, \\
m &= 10^{-7}\,\mathrm{kg}, \\
F &= 4\times 10^5,\\
\lambda_{L} &= 532\,\mathrm{nm},
\end{align*}
we obtain $(\Delta \beta_{0})^{2} = 10^{18}$ ($10^{24}$) in the quantum-noise-limited (classical-noise-limited) scheme.
Thus, by performing $10^{4}$ runs of the quantum-noise-limited experiment, an upper bound of $\beta + \Delta \beta < 10^{7}$ can be attained, which is still 26 orders of magnitude better than present bounds.
As described in~Appendix~\ref{App:Distortion}, the state undergoes possible distortion because of the nonlinear (in $n$) terms in the unitary operator, but this distortion is expected to be insignificant for current experimental parameters. 
Assuming that the distortion does not significantly affect the phase statistics, we see that if we also have squeezing with $r=3$, we get $(\Delta \beta_{0})^{2} = 10^{15}$ ($10^{21}$).

We now calculate the number of runs if we use the four-loop path. Performing similar calculations, we get number of runs to be $N_{r} = 10^{16}$ ($10^{22}$). As expected, the precision is higher in this case but the accuracy is possibly lower because of the uncontrolled approximation. Using a squeezing parameter $r = 3$, we can further reduce the number of runs by three orders of magnitude; $N_{r} = 10^{13}$ ($10^{19}$).

\subsubsection{$\mu_{0}$ case}
\label{Sec:MuResult}

\begin{figure}[h]\centering
 \includegraphics[height=0.3\columnwidth]{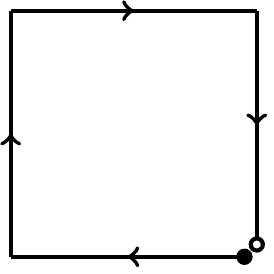}
\caption{$U_{\mu_{0}}$: The shape of the path to reduce QM contribution for the $\mu$ commutator.}
\label{Fig:UMu}
\end{figure}

In the $\mu_{0}$ case, the largest quantum mechanical term cannot be removed from the total phase. 
So, the path in phase space is just a rectangular loop like the original path~\cite{Pikovski2012}.
However, choosing a different starting point leads to smaller quantum mechanical terms in total.
The optimal path is effected by the unitary operator
\begin{equation}
U_{\mu_{0}} =
\e^{-\iota H_{X}} 
\e^{-\iota H_{P}} 
\e^{\iota H_{X}}
\e^{\iota H_{P}}
\end{equation}
and is depicted in Figure~\ref{Fig:UMu}.

Performing calculations similar to the $\gamma_{0}$ and $\beta_{0}$ cases, the total phase $\Phi_{T}$ is given by 
\begin{equation}
\Phi_{T} = 2 \mu \lambda_{0}^{2} N_{p} 
+ 2 \lambda_{0}^{2} N_{p} 
- 2k^{3} \lambda_{0}^{5} N_{p}^{4}
+ 35 k^{4} \lambda_{0}^{6} N_{p}^{5}.
\label{Eq:AppPhiTMu0}
\end{equation}
Here, as in the second $\beta_{0}$ case, the largest contribution to the quantum mechanical phase remains the same. 
However, choosing a different starting point in the loop reduces the second-largest term by seven orders of magnitude. 
This leads to a marginal improvement in the number of runs with no extra experimental cost.

From Equation~\eqref{Eq:AppPhiTMu0}, the value of $\mu_{0}$ is estimated as
\begin{equation}
\mu_0 = \frac{1}{ 2 \kappa'' \lambda_0^2} \left( \frac{\Phi_T}{N_p}\right) 
- \frac{1}{\kappa''}
+ \frac{ k^{3} \lambda_{0}^{3} N_{p}^{3}} {\kappa''}
- \frac{35 k^{4} \lambda_{0}^{4} N_{p}^{4}} {2 \kappa''}
\end{equation}
for 
\begin{equation}
\kappa'' := \frac{m^{2}}{M_{p}^{2}}.
\end{equation}
The variance in $\mu_{0}$ is therefore given by
\begin{align}
& \left( \Delta \mu_0 \right)^2 =
\left( \frac{1}{ 2 \kappa'' \lambda_0^2 N_p} \right)^2 \left(\Delta \Phi_{T} \right)^2 
+ \left(- \frac{ \Phi_T}{ 2 \kappa'' \lambda_0^2 N_p^2} 
\right. \nonumber \\
& \quad \left.
+ \frac{3 k^3 \lambda_{0}^3}{\kappa''} N_p^2 
- \frac{70 k^4 \lambda_{0}^4}{\kappa''} N_p^3 
\right)^2 \left( \Delta N_p \right)^2 
\label{Eq:MuUncertainty}
\end{align}
for one run of the experiment.
 
We evaluate the expression first for coherent states. 
For the same experimental parameters in the original proposal, we evaluate $(\Delta \mu_{0})^{2} = 10^{5}$ in both cases, which is the same as before because the path is almost the same.
However we note that the error decreases monotonically with $N_{p}$, with $m$ and with $F$, hence the highest possible value of these parameters should be chosen for the experiment.
Keeping the parameters
\begin{align}
\lambda_L &=1064\,\mathrm{nm}\\
L &= 4 \,\mu\mathrm{m}\\
\omega_m &= 2\pi \times 10^5\\
F &= 10^5
\end{align}
fixed and changing the mean photon number and the oscillator mass to
\begin{align}
N_p &= 10^{9}, \\
m &= 10^{-10}\,\mathrm{kg}, 
\end{align}
we obtain $(\Delta \mu_{0})^{2} = 2.2$ (22) for a single run of the experiment in the quantum-noise-limited (classical-noise-limited) scheme.
If we also include using squeezed light with the squeezing parameter $r=-3$, the variance in the quantum-noise-limited scheme further reduces to $(\Delta \mu_{0})^{2} = 10^{-3}$.
We can further increase the signal to noise ratio by increasing the mass of the oscillator.


%

\end{document}